\documentclass[preprint,onecolumn,nofootinbib]{revtex4}
\usepackage[colorlinks=true,linkcolor=blue,urlcolor=blue,filecolor=black,citecolor=red,pdfstartview=FitV,pdftitle={},pdfsubject={},pdfkeywords={},pdfpagemode=None,bookmarksopen=true]{hyperref}
\usepackage{graphicx}
\usepackage{amsmath}
\usepackage{amsfonts}
\usepackage{amssymb,ulem}
\usepackage{color,xcolor}
\usepackage{CJK}
\usepackage{subfigure}
\usepackage{amsthm,amsmath,amssymb}
\usepackage{mathrsfs}
\usepackage{url}
\usepackage{hyperref}
\usepackage{array}
\usepackage{dcolumn}
\usepackage{float}
\usepackage{appendix}
\usepackage{subcaption}
\begin{document}
\title{Gravitational waveforms from periodic orbits around a novel regular black hole}

\author{Huajie Gong$^{1}$}
\thanks{huajiegong@hunnu.edu.cn}
\author{Sheng Long$^{2}$}
\author{Xi-Jing Wang$^{3}$}
\author{Zhongwu Xia$^{1}$}
\author{Jian-Pin Wu$^{4}$}
\thanks{jianpinwu@yzu.edu.cn, corresponding author}
\author{Qiyuan Pan$^{1,4}$}
\thanks{panqiyuan@hunnu.edu.cn, corresponding author}

\affiliation{
	$^{1}$ \mbox{Department of Physics, Institute of Interdisciplinary Studies, Key Laboratory of} \mbox{Low Dimensional Quantum Structures and Quantum Control of Ministry of Education,} \mbox{Synergetic Innovation Center for Quantum Effects and Applications, and Hunan Research} \mbox{Center of the Basic Discipline for Quantum Effects and Quantum Technologies,} \mbox{Hunan Normal University, Changsha 410081, China}\\
    $^{2}$ \mbox{School of Fundamental Physics and Mathematical Sciences, Hangzhou Institute for}
    \mbox{Advanced Study, University of Chinese Academy of Sciences, Hangzhou 310024, China}\\
	$^{3}$ \mbox{School of Physics and Technology,} \mbox{Wuhan University, Wuhan 430072, China}\\
    $^{4}$ \mbox{Center for Gravitation and Cosmology, College of Physical Science and Technology,} \mbox{Yangzhou University, Yangzhou 225009, China}\\
}

\begin{abstract}
We explore potential quantum gravity signatures by studying periodic orbits and their GW emissions around a novel regular black hole (BH) featuring a Minkowski core. Using a rational number $q$, periodic orbits are classified, revealing that the deviation parameter $\alpha_0$ reshapes the bound-orbit region while preserving characteristic ``zoom-whirl" structures. Numerical kludge waveforms reveal detectable phase shifts and amplitude modulations induced by quantum gravity effects with radiation reaction breaking orbital periodicity. Faithfulness analysis demonstrates that larger $\alpha_{0}$ and $q$ enhance distinguishability from the Schwarzschild case, and a comparison with Hayward and quantum  Oppenheimer-Snyder BHs shows their similar large-scale behaviors yield macroscopically indistinguishable orbits and waveforms.

\end{abstract}

\maketitle
\tableofcontents

\section{Introduction}\label{sec:intro}

The groundbreaking advancements by the LIGO and Virgo collaborations in gravitational wave (GW) detection have provided robust confirmation of the predictions of general relativity (GR) in the strong-field regime \cite{LIGOScientific:2016vpg,LIGOScientific:2016aoc,LIGOScientific:2021usb,LIGOScientific:2020ibl,KAGRA:2021vkt}. With the continuous improvement of detector sensitivities, particularly the forthcoming deployment of space-based GW detectors such as LISA \cite{LISA:2017pwj}, Taiji \cite{Hu:2017mde} and TianQin \cite{TianQin:2015yph}, it is becoming increasingly feasible to use GW signals to probe physics beyond GR. Among the most actively pursued directions is the search for potential observational signatures of quantum gravity effects, including GW echoes \cite{Cardoso:2016oxy,Abedi:2016hgu,Tan:2024qij}, deviations in the geometry of black hole (BH) shadows \cite{EventHorizonTelescope:2019dse,EventHorizonTelescope:2020qrl}, and subtle phase shifts in the waveforms of the extreme mass-ratio inspiral (EMRI) systems \cite{Barack:2018yly}.

The Penrose-Hawking singularity theorems demonstrate that BHs inevitably encounter the singularity problem within the framework of GR, suggesting that quantum gravity effects may dominate in the strong-curvature regime \cite{Penrose:1964wq,Hawking:1973uf}. Among the various proposals aimed at resolving this issue, regular black holes (RBHs), referring to BH solutions that avoid curvature singularities at the center, have attracted considerable attention in recent years \cite{Dymnikova:1992ux,Bronnikov:2005gm,Ansoldi:2008jw,Fan:2016hvf,Cisterna:2020rkc,Babichev:2020qpr,Bronnikov:2021uta,Bokulic:2022cyk,Canate:2022zst,Chew:2022enh,Barrientos:2022avi,Capozziello:2024ucm,Capozziello:2025wwl}. Such solutions are typically constructed in a phenomenological manner, most of which can be realized by introducing exotic matter that violate classical energy conditions, for instance, through nonlinear electrodynamics sources \cite{Ayon-Beato:1998hmi,Bronnikov:2000vy}. Moreover, RBHs are often considered potential manifestations of quantum gravitational effects. Based on their asymptotic behavior near the center, RBHs can be classified into two categories: those featuring a de-Sitter core \cite{Bardeen:1968ebo,Hayward:2005gi,Frolov:2014jva} and those possessing a Minkowski core \cite{Xiang:2013sza,Culetu:2013fsa,Balart:2014cga,Simpson:2019mud}. Recently, a novel class of Minkowski-core RBHs has been proposed whose Kretschmann scalar remains sub-Planckian throughout the entire spacetime. This property enables the model to apply to BHs of any mass and to exhibit physically more reasonable behavior toward the end of the evaporation process \cite{Ling:2021olm,LingLingYi:2021rfn}. Subsequent studies have further explored features of this class of BHs, including the extremal stable circular orbits \cite{Zeng:2022yrm}, shadow and image structures \cite{Ling:2022vrv,Zeng:2023fqy}, and quasi-normal modes \cite{Yang:2023gas,Xia:2023zlf,Zhang:2024nny,Yang:2024ofe}, highlighting their particular appeal for investigating quantum gravity effects through astronomical observations.

The timelike geodesics of massive particles around a BH provide a powerful probe of the spacetime geometry. Within the Newtonian framework, the motion of such particles can be classified into elliptical, parabolic, or hyperbolic trajectories. In GR, however, the dynamics become considerably more intricate. For example, in an EMRI system, the motion of a stellar-mass BH, well approximated as a test particle, orbiting a supermassive BH has  orbital precession, resulting in distinctive relativistic features \cite{Yang:2017aht,Cui:2025bgu,Han:2018hby,Zi:2021pdp,Zhang:2022rfr,Rahman:2023sof,Qiao:2024gfb, Tan:2024hzw,Fu:2024cfk,Zi:2024jla,Kumar:2024our,Xia:2025yzg}. During the gradual inspiral of the EMRI, periodic orbits can serve as continuously evolving transition states. These orbits are expected to facilitate modeling of adiabatic EMRI waveform \cite{Glampedakis:2002ya}.

Periodic orbits, as a special subclass of bound trajectories, contain fundamental information about the orbital dynamics around a central BH \cite{Poisson:2003nc,Levin:2008mq}. Owing to the strong curvature of BH spacetimes, the motion of the particle generally exhibits intricate quasi-periodic behavior. However, under specific conditions of the energy and angular momentum, the motion of a small-mass object around a supermassive BH can be characterized by a rational ratio between its radial and angular frequencies. 
Such configurations are associated with orbital resonances which are expected to play a significant role in GW physics \cite{Healy:2009zm,Apostolatos:2009vu,Flanagan:2010cd,Berry:2016bit,Speri:2021psr}. In these cases, the accumulated azimuthal angle over one period is exactly an integer multiple of $2\pi$, leading to a closed periodic orbit. Following the classification scheme proposed in Ref.~\cite{Levin:2008mq}, each periodic orbit can be uniquely characterized by a triplet $(z, w, v)$, where $z$, $w$, and $v$ describe the orbit’s zoom, whirl and vertex behaviors, respectively. These orbits oscillate between periapsis and apoapsis, tracing distinctive spatial patterns that provide an important window into the spacetime structure in the strong gravitational regime. The classification method has been applied to particle motion in a wide variety of BH spacetimes, including Schwarzschild BHs \cite{Levin:2008mq, Lim:2024mkb}, Kerr BHs \cite{Levin:2008ci, Levin:2008yp, Levin:2009sk, Grossman:2011ps}, charged BHs \cite{Misra:2010pu}, quantum-corrected BHs \cite{Deng:2020yfm, Tu:2023xab, Yang:2024lmj} and many other background spacetimes \cite{Liu:2018vea, Babar:2017gsg, Azreg-Ainou:2020bfl, Wei:2019zdf, Zhou:2020zys, Lin:2021noq, Gao:2021arw, Zhang:2022zox, Wang:2022tfo, Zhang:2022psr, Lin:2023rmo, Yao:2023ziq, Huang:2024oli, Jiang:2024cpe, Li:2024tld, Meng:2024cnq, Zhao:2024exh, QiQi:2024dwc,Junior:2024tmi, Shabbir:2025kqh,Chan:2025ocy,Chen:2025ncm,Guo:2025scs}.

The paper is organized as follows. In Section~\ref{sec:ds2andEOM}, we provide a brief review of the novel RBH spacetime with a Minkowski core and analyze the orbital dynamics of a test particle in this spacetime. Section~\ref{sec:periodicorbits} presents the classification of periodic orbits in the RBH. In Section~\ref{sec:NKwaveforms}, we outline the numerical kludge approach and generate gravitational waveforms from representative periodic trajectories. In Section~\ref{sec:reular-Hay-LQG}, we compare the dynamical and waveform characteristics of the novel RBH with Hayward BHs and quantum  Oppenheimer-Snyder (qOS) BHs. Finally, in Section~\ref{sec:conclusion}, we summarize our main results and suggest possible avenues for future research. Appendix~\ref{ceffect} examines the influence of the parameter $c$ in the novel RBH metric on orbital dynamics and GW waveforms, while Appendix~\ref{erroranalysis} conducts a numerical error analysis in this work.

\section{Novel regular black holes (RBHs) and the test particle dynamics}\label{sec:ds2andEOM}
\subsection{The novel RBH spacetime}\label{subsec:ds2}
The earliest known model of a RBH with a Minkowski core was first constructed phenomenologically in Ref.~\cite{Xiang:2013sza}, motivated by the aim of resolving the central singularity of the Schwarzschild BH. The metric is given by
\begin{eqnarray} \label{metric0}
	ds^2=-(1+2 \varphi(r))dt^2+\frac{1}{1+2 \varphi(r)}dr^2+ r^2 d \theta^2 + r^2 \sin^2 \theta d \phi^2\,,~~~ \varphi(r)=-\frac{M}{r}e^{\frac{-\tilde{\alpha}}{r^2}},
\end{eqnarray}
where $\varphi(r)$ represents the modified gravitational potential, and the constant $\tilde{\alpha}$ is assumed to be of the order of the square of the Planck length\footnote{Note that in this work, we set $c = G = \hbar = 1$, so the Planck length is ${l_{\text{P}}} = 1$.}. Evidently, the exponential suppression causes $\varphi(r)$ to vanish as $r \to 0$, thereby replacing the Schwarzschild singularity with an asymptotically Minkowski region near the center. The parameter $\tilde{\alpha}$ is closely related to quantum gravitational effects, which can be understood from the following two perspectives:
\begin{itemize}
		\item In quantum gravity, it is widely believed that there exists a minimal observable length, typically on the order of the Planck scale. This idea is naturally incorporated in the Generalized Uncertainty Principle (GUP)~\cite{Harbach:2005yu}. As mentioned in the article~\cite{Harbach:2005yu}, the GUP form proposed by Harbach and Hossenfelder is:
		\begin{eqnarray}
			[p, x] = -i e^{\alpha' p^2}\,,
		\end{eqnarray}
		where $\alpha'$ is related to the minimal length scale. Through a heuristic argument, the authors show that GUP leads to a running Newton gravitational constant: $G = G_N / z(p)$. In the weak-field approximation, this yields a modified gravitational potential whose form matches $\varphi(r)$ in the metric function (\ref{metric0}), establishing a close relation between $\tilde{\alpha}$ and $\alpha'$. Thus, $\tilde{\alpha}$ can be regarded as the manifestation of the minimal length scale introduced by GUP in the BH metric.
		\item From the perspective of the Einstein field equations, the modified metric can be attributed to an effective matter fluid (see Eq.~(14) in the original paper~\cite{Xiang:2013sza}). This effective matter field violates the strong energy condition in the region $r < \sqrt{2\tilde{\alpha}/3}$. This is seen as a signature of quantum gravitational effects, since classical matter usually satisfies energy conditions, whereas quantum effects may lead to their violation.
\end{itemize}

Following the above work, many RBH models based on different forms of such exponentially suppressing potentials have been proposed and discussed \cite{Culetu:2013fsa,Culetu:2014lca,Balart:2014cga,Ghosh:2014pba,Ghosh:2018bxg,Rodrigues:2015ayd,Simpson:2019mud,Ling:2021olm,LingLingYi:2021rfn}. In this work, we adopt the metric of a novel class of RBHs model introduced in Refs.~\cite{Ling:2021olm,LingLingYi:2021rfn}, which takes the form:
\begin{eqnarray} \label{metric}
	ds^2=-f(r)dt^2+\frac{1}{f(r)}dr^2+ r^2 d \theta^2 + r^2 \sin^2 \theta d \phi^2\,,
\end{eqnarray}
where
\begin{eqnarray} \label{fr}
	f(r)=1+2 P(r)\,,~~~P(r)=-\frac{M}{r}e^{\frac{-\alpha_{0}M^x}{r^c}}\,.
\end{eqnarray}
Here, $P(r)$ represents the modified Newtonian potential with the parameter $\alpha_0$ characterizing the deviation from the classical Newtonian form\footnote{Note that $\alpha_{0}$ here is analogous to $\tilde{\alpha}$ in \cite{Xiang:2013sza}.}. The symbol $e$ in $P(r)$ denotes the base of the natural logarithm, commonly known as Euler's number. Furthermore, we emphasize that the exponential form $\exp(-\alpha_0 M^x/r^c)$ is constructed phenomenologically to ensure a smooth decay of the spacetime curvature toward the center, keeping it below the Planck scale. This feature is essential for resolving the classical singularity, in line with motivations from quantum gravity. When $\alpha_0 = 0$, the metric function $f(r)$ reduces to that of the Schwarzschild solution. The quantities $x$ and $c$ are additional dimensionless parameters\footnote{The study on the effects of the parameter $c$ is provided in the Appendix~\ref{ceffect}.}. Throughout this work, we fix $x = 1$ and $c = 3$, under which this novel RBH exhibits large-scale behavior similar to that of the Hayward BH~\cite{Hayward:2005gi} and the quantum Oppenheimer-Snyder (qOS) BH \cite{Lewandowski:2022zce}.

\begin{figure}[htbp]
	\centering
	\includegraphics[width=0.48\textwidth]{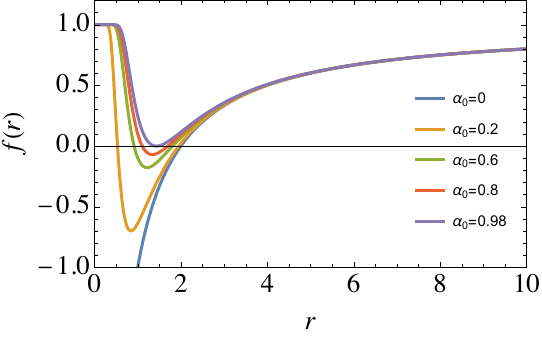}\hspace{4mm}
	\includegraphics[width=0.48\textwidth]{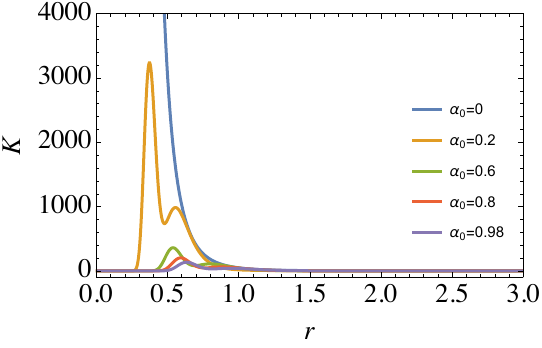}\hspace{4mm}
	\caption{Left panel: The metric function $f(r)$ for different deviation parameters $\alpha_{0}$. Right panel: The Kretschmann scalar $K$ for different deviation parameters $\alpha_{0}$.}
	\label{FigfrandK}
\end{figure}

As shown in Ref.~\cite{Zhang:2024nny}, the requirement of the event horizon imposes a constraint on $\alpha_0$, restricting it to the interval $0\leqslant \alpha_0\leqslant 8M^2/3e$. Note that the parameter $\alpha_{0}$ has the dimension of the square of the length. Without loss of generality, the RBH mass is set to $M=1$. To aid visualization, we plot the metric function $f(r)$ in the left panel of Fig.~\ref{FigfrandK}. One observes that as $\alpha_0$ increases, the curves exhibit more pronounced variation in the small-radius region. This behavior arises because a larger $\alpha_0$ leads to a faster decay rate of the exponential term.

To analyze the properties of this RBH spacetime, we explicitly compute the Kretschmann scalar $K$ as
\begin{eqnarray} \label{Kexpression}
	K=R^{\mu\nu\rho\sigma}R_{\mu\nu\rho\sigma}=\frac{12 e^{-\frac{2 \alpha_0}{r^3}} \left(27	\alpha_0^4+4 r^{12}-32 \alpha_0 r^9+132 \alpha_0^2 r^6-108 \alpha_0^3 r^3\right)}{r^{18}}.
\end{eqnarray}
The behavior of $K(r)$ is shown in the right panel of Fig.~\ref{FigfrandK} for several values of $\alpha_0$. As illustrated, in the Schwarzschild case ($\alpha_0=0$), $K(r)$ diverges as $r\to 0$ and decreases rapidly toward zero as $r$ increases. For nonzero $\alpha_0>0$, the divergence is eliminated, and the overall amplitude of $K(r)$ decreases monotonically with increasing $\alpha_0$. In particular, one observes that as $\alpha_0$ increases, the curves exhibit more pronounced variation in the small-radius region. This behavior arises because a larger $\alpha_0$ increases the decay rate of the exponential term. Notably, for all $\alpha_0>0$, $K(r)$ tends smoothly to zero both as $r\to 0$ and as $r\to \infty$, confirming the absence of a central singularity and the recovery of asymptotic flatness.

These asymptotic behaviors can also be derived analytically from Eq.~\eqref{Kexpression}. In the limit $r \to 0$, we find $K(r) \sim e^{-2\alpha_0/r^3} r^{-18} \to 0$, where the exponential factor dominates and ensures that the curvature vanishes smoothly at the center. In the asymptotic region $r \to \infty$, we obtain $K(r) \approx \frac{48M^2}{r^6} -\frac{480 \alpha_0}{r^9}+ \mathcal{O}(1/r^{10})$, which reproduces the Schwarzschild behavior, with $\alpha_0$-dependent corrections appearing at higher orders. These analytical results are fully consistent with the numerical behavior displayed in the right panel of Fig.~\ref{FigfrandK}.

\subsection{Dynamics of massive test particles}
We now consider the motion of a massive test particle in this RBH spacetime. The particle dynamics are governed by the Lagrangian \cite{Chandrasekhar:1985kt}
\begin{eqnarray} \label{Lagrangian}
\mathscr{L} = \frac{1}{2} g_{\mu \nu} \dot{x}^\mu \dot{x}^\nu,
\end{eqnarray}
where $g_{\mu \nu}$ is the background metric and the overdot on the $x$ denotes the derivative with respect to the proper time. For a massive particle, it must satisfy the normalization condition $\mathscr{L} = -1/2$. From this Lagrangian, the four-momentum of the particle is
\begin{eqnarray} \label{momentum}
p_{\mu} = \frac{\partial \mathscr{L}}{\partial \dot{x}^\mu} = g_{\mu \nu} \dot{x}^\nu.
\end{eqnarray}
Given the static and spherically symmetric nature of the spacetime~\eqref{metric}, the system possesses two conserved quantities: the energy $E$ and the angular momentum $L$. Substituting Eqs. \eqref{metric} and \eqref{Lagrangian} into Eq. \eqref{momentum}, one can derive the equations of motion for the particle:
\begin{eqnarray} \label{fourv}
	p_t &=& g_{tt} \dot t  = - E, \label{pt}\\
	p_\phi &=& g_{\phi \phi} \dot \phi = L,  \label{pphi}\\
	p_r &=& g_{rr} \dot r,\\
	p_\theta &=& g_{\theta \theta} \dot \theta.
\end{eqnarray}
By solving Eqs. \eqref{pt} and \eqref{pphi}, we obtain the expressions for $\dot{t}$ and $\dot{\phi}$ in terms of the conserved quantities $E$ and $L$:
\begin{eqnarray}
	\dot t = - \frac{ E  }{ g_{tt} } = \frac{E}{f(r)},\,\,\,\,\,\,\,\,\,\,	\dot \phi = \frac{L}{g_{\phi\phi}} = \frac{L}{r^2 \sin^2\theta}. \label{tphidot}
\end{eqnarray}
Due to the spherical symmetry of the spacetime \eqref{metric}, we focus on the particle moving on the equatorial plane ($\theta = \pi/2$) for convenience.
Using the normalization condition $g_{\mu \nu} \dot x^\mu \dot x^\nu = -1$ together with Eq.~\eqref{tphidot}, we obtain
\begin{eqnarray}\label{EOM}
	\left(-1+\frac{2 M e^{-\alpha_{0}M/r^{3}}}{r}\right) \dot t^2 + \left(1+\frac{2}{e^{\alpha_{0}M/r^{3}}r-2}\right) \dot r^2 +  r^2\dot \phi^2 = -1.
\end{eqnarray}
This can be rewritten as
\begin{eqnarray}\label{GeodesicV2}
	\dot r ^2 + V_{\rm eff}(r)= E^2\,,
\end{eqnarray}
where the effective potential is given by
\begin{eqnarray}\label{Veff}
V_{\rm eff}(r)= \left(1-\frac{2 M e^{-\alpha_{0}/r^{3}}}{r}\right)\left(1+\frac{L^2}{r^2} \right)\,.
\end{eqnarray}

\begin{figure}[htbp]
	\centering
	\includegraphics[width=0.48\textwidth]{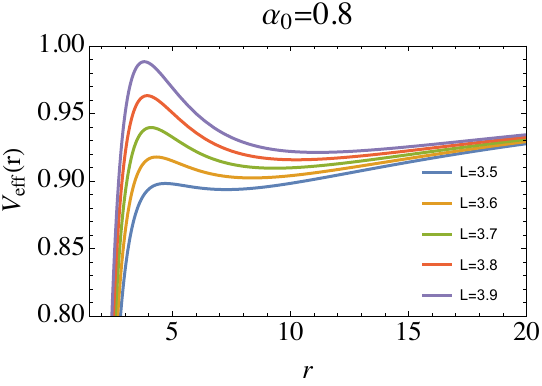}\hspace{4mm}
	\includegraphics[width=0.48\textwidth]{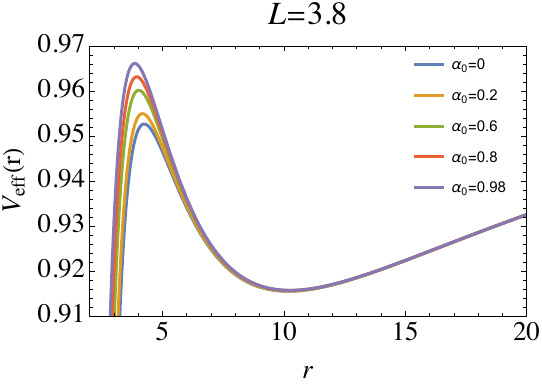}\vspace{0.4mm}
	\caption{Left panel: The effective potential $V_{\text{eff}}$ for different orbital angular momentum $L$ with the deviation parameter $\alpha_{0}=0.8$. Right panel: The effective potential $V_{\text{eff}}$ for different deviation parameters $\alpha_{0}$ with orbital angular momentum $L=3.8$.}
	\label{Veffv1}
\end{figure}

Without loss of generality, we set the maximum energy as $E = 1$. From the expression of the effective potential (Eq.~\eqref{Veff}), it follows that $V_{\rm eff} \to 1$ as $r \to \infty$. According to the same equation, if the energy of the particle is $E>1$, it can escape to infinity, implying an unbound orbit. For bound orbits, Fig.~\ref{Veffv1} illustrates the behavior of the effective potential $V_{\rm eff}$ for different values of orbital angular momentum $L$ and the deviation parameter $\alpha_0$. As shown in the left panel, a local maximum emerges in the effective potential, growing in magnitude with increasing $L$. Similarly, the right panel shows the effective potential $V_{\rm eff}$ exhibits a local maximum whose values increase with $\alpha_0$. These results indicate that increasing either $L$ or $\alpha_0$ raises the upper energy limit for which particles can have bound orbits. In other words, higher angular momentum or stronger deviation from the Schwarzschild metric increases the maximum energy at which bound motion is still possible.

\begin{figure}[htbp]
	\centering
	\includegraphics[width=0.48\textwidth]{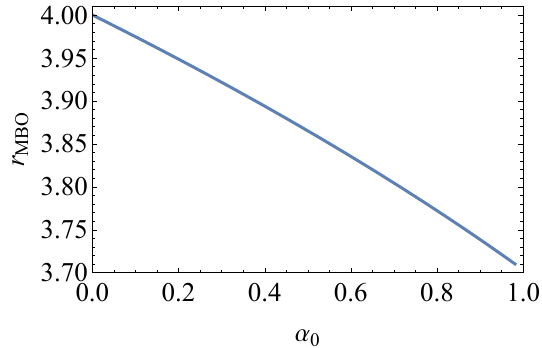}\hspace{4mm}
	\includegraphics[width=0.48\textwidth]{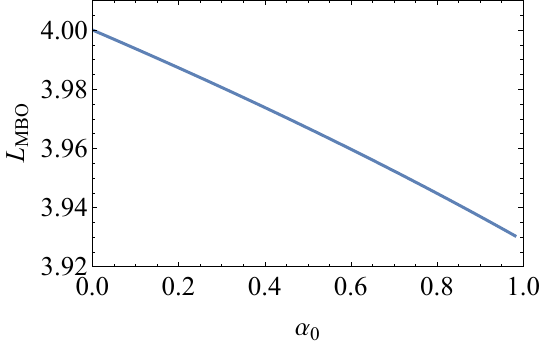}\vspace{0.4mm}
	\caption{The radius $r_\text{MBO}$ (left panel) and the angular momentum $L_\text{MBO}$ (right panel) as a function of the deviation parameter $\alpha_{0}$.}
	\label{MBOvsa0}
\end{figure}
Next, we investigate the properties of bound orbits in this RBH spacetime. Generally, periodic orbits are a special class of bound orbits which exist between the marginal bound orbit (MBO) and the innermost stable circular orbit (ISCO). The behavior of a particle in bound orbits can be characterized by the extremum of the effective potential $V_{\text{eff}}$. The MBO satisfies the following conditions
\begin{eqnarray}
	V_{\text{eff}}=1,~~~\frac{dV_{\text{eff}}}{dr}=0,
\end{eqnarray}
which correspond to an unstable circular orbit at the maximum energy required for the particle to escape. Fig.~\ref{MBOvsa0} indicates that, compared to the Schwarzschild BH in GR, the parameter $\alpha_{0}$ leads to a gradual decrease in both the radius and angular momentum of the MBO. This implies that quantum gravity effects allow particles to remain bound to the BH at closer distances, while requiring lower angular momentum to maintain orbital stability.

\begin{figure}[htbp]
	\centering
	\includegraphics[width=0.48\textwidth]{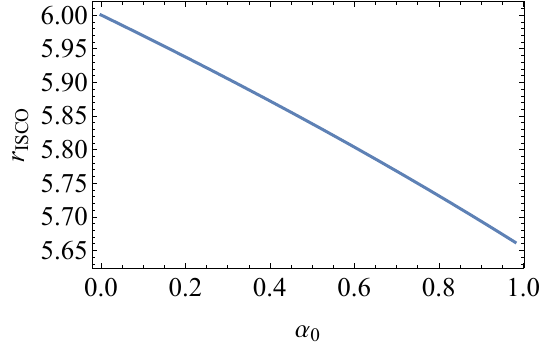}\hspace{4mm}
	\includegraphics[width=0.48\textwidth]{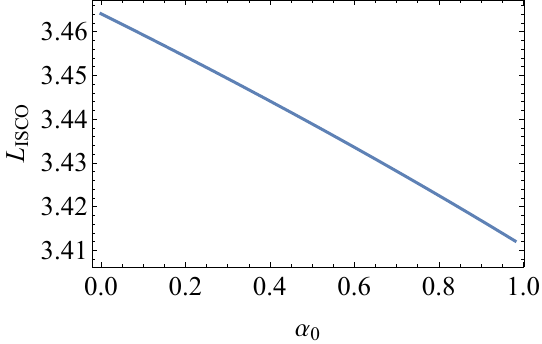}\vspace{0.4mm}
	\includegraphics[width=0.48\textwidth]{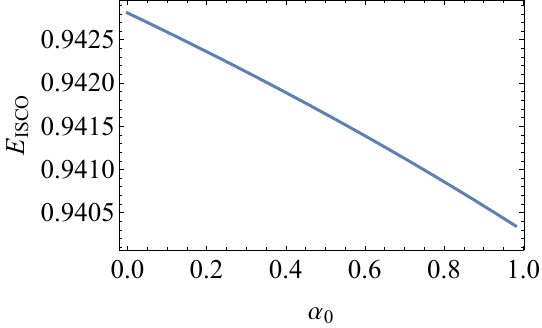}\vspace{0.4mm}
	\caption{The radius $r_\text{ISCO}$ (upper left panel), the angular momentum $L_\text{ISCO}$ (upper right panel) and the energy $E_\text{ISCO}$ (lower panel) as a function of the deviation parameter $\alpha_{0}$.}
	\label{ISCOvsa0}
\end{figure}
On the other hand, the ISCO represents the smallest orbital radius at which a particle can stably orbit around the BH, determined by the following three conditions simultaneously
\begin{eqnarray}
	V_{\text{eff}}=E^2,~~~\frac{dV_{\text{eff}}}{dr}=0,~~~\frac{d^2V_{\text{eff}}}{dr^2}=0.
\end{eqnarray}
It is evident that the conditions for the ISCO are more stringent than those for the MBO. Fig.~\ref{ISCOvsa0} shows that as the deviation parameter $\alpha_0$ increases, the radius, energy, and angular momentum of the ISCO all exhibit a monotonically decreasing trend. In particular, the variation of the ISCO radius indicates that, for a given $\alpha_0$, the ISCO in this RBH spacetime can be located closer to the central object than in the Schwarzschild spacetime.

\begin{figure}[htbp]
	\centering
	\includegraphics[width=0.6\textwidth]{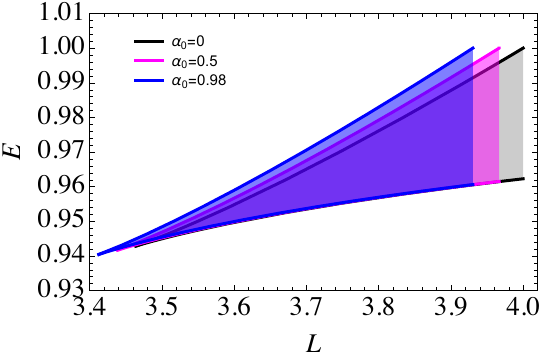}\hspace{4mm}
	\caption{The allowed $(E-L)$ regions for the massive particle’s bound orbits with different deviation parameters $\alpha_{0}$.}
	\label{LErange}
\end{figure}
In Fig.~\ref{LErange}, we plot the allowed region of the energy $E$ and angular momentum $L$ required for the formation of bound orbits in the RBH spacetime, where the gray region corresponds to the case of the Schwarzschild BH. Obviously, as the deviation parameter $\alpha_{0}$ increases, the allowed region for bound orbits shifts relative to the Schwarzschild case.

\section{Periodic orbits}\label{sec:periodicorbits}

In this section, we investigate a special class of bound orbits—periodic orbits—in the spherically symmetric RBH spacetime. A key feature of such orbits is that a timelike particle following the trajectory returns precisely to its initial position after a finite evolution time. According to Ref. \cite{Levin:2008mq}, any generic (non-periodic) orbit can be approximated by a nearby periodic orbit; equivalently, a generic orbit may be viewed as a perturbation of some underlying periodic motion. Therefore, the study of periodic orbits is of significant importance for understanding the structure of general bound trajectories and their associated GW signatures, playing a crucial role in GW physics.

In the spherically symmetric spacetimes considered here, it is shown in Ref. \cite{Levin:2008mq} that an orbit is periodic if and only if the ratio of its angular frequency $\omega_\phi$ to its radial frequency $\omega_r$ is a rational number. Such orbits can be characterized by a rational number $q$, or equivalently by a triplet of integers $(z, w, v)$, defined as
\begin{eqnarray}\label{DefineQ}
	q\equiv w + \frac{v}{z}=\frac{\omega_\phi}{\omega_r}-1 =\frac{\bigtriangleup\phi}{2\pi}-1,
\end{eqnarray}
where the triplet $(z, w, v)$ characterizes the zoom, whirl, and vertex behaviors of a periodic orbit, respectively. To avoid orbital degeneracy \footnote{This degeneracy refers to the fact that for a given value of $w$, multiple $(z, v)$ pairs may correspond to the same physical orbit.}, the integers $z$ and $v$ are required to be relatively prime. The angular and radial frequencies, $\omega_\phi$ and $\omega_r$, which govern the motion in the $\phi$ and $r$ directions, respectively, are defined as follows:
\begin{eqnarray}
	\omega_r=\frac{2\pi}{T_r},~~~~~~\omega_\phi=\frac{1}{T_r} \int_0^{T_r}\frac{d\phi}{dt} dt =\frac{\Delta\phi}{T_r},
\end{eqnarray} 
where $T_r$ denotes the (coordinate) time required for a particle to complete one full radial cycle, while $\Delta \phi$ represents the total accumulated angle swept by the particle during a single radial period. The latter is given by the following expressions:
\begin{eqnarray}
	\bigtriangleup\phi=2\int_{r_a}^{r_p}\frac{\dot \phi}{\dot r} dr=2 \int_{r_a}^{r_p} \frac{L}{r^2\sqrt{E^2-V_{\text{eff}}(r)}}dr.
\end{eqnarray}
Here, $r_a$ and $r_p$ are the turning points known as the apastron and periastron, respectively, which lie between the ISCO and the MBO. These can be obtained by solving Eq.~\eqref{GeodesicV2} numerically.

\begin{figure}[htbp]
	\centering
	\includegraphics[width=0.48\textwidth]{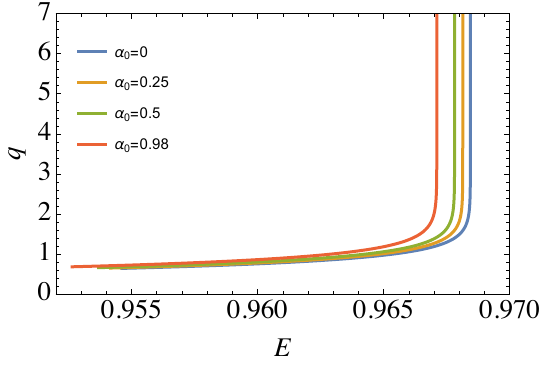}\hspace{4mm}
	\includegraphics[width=0.46\textwidth]{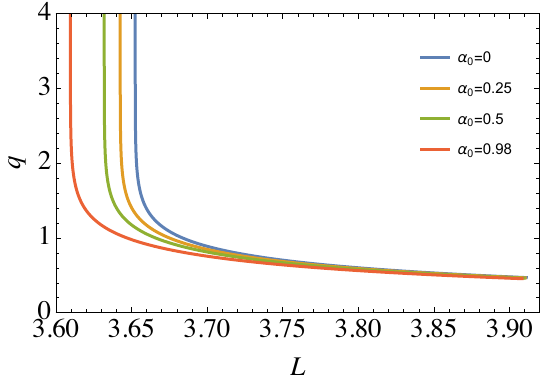}\vspace{0.4mm}
	\caption{Left panel: The rational number $q$ as a function of the energy of periodic orbits $E$ for the RBH with different values of the parameter $\alpha_0$. The orbital angular momentum is fixed as $L = (L_\text{MBO} + L_\text{ISCO})/2$. Right panel: The rational number $q$ as a function of the angular momentum of periodic orbits $L$ for the RBH with different values of the parameter $\alpha_0$. The energy is fixed as $E = 0.96$.}
	\label{qvsEandLv1}
\end{figure}
In Fig.~\ref{qvsEandLv1}, we illustrate how the rational number $q$ varies with the energy $E$ and the angular momentum $L$ under different values of the parameter $\alpha_0$.  In the left panel, where $L$ is held constant, $q$ initially increases slowly with the energy, followed by a sharp rise as $E$ approaches its maximum value. For a fixed value of $q$, it is evident that a larger $\alpha_0$ corresponds to a lower maximum energy required to sustain the orbit. In the right panel , with the energy $E$ fixed instead, we observe that $q$ generally decreases as angular momentum increases, but diverges to positive infinity as $L$ approaches its minimum value. Likewise, for a given $q$, an increase in $\alpha_0$ leads to a reduction in the minimum angular momentum required to sustain the orbit.

\begin{table}[htbp]
	\centering
	\setlength\tabcolsep{6pt}
	\caption{The energy $E$ for the periodic orbits with different values of $(z, w, v)$ and the deviation parameter $\alpha_0$. The orbital angular momentum is fixed as $L = (L_\text{MBO} + L_\text{ISCO})/2$.}
	\begin{tabular}{|c|c|c|c|c|c|c|c|c|c|c|c|c|}
		\hline
		$\alpha_0$ & $E_{(1,1,0)}$ & $E_{(1,2,0)}$ & $E_{(2,1,1)}$ & $E_{(2,2,1)}$ & $E_{(3,1,2)}$ & $E_{(3,2,2)}$ & $E_{(4,1,3)}$ & $E_{(4,2,3)}$\\
		\hline
		0 & 0.965425 & 0.968383 & 0.968026 & 0.968434 & 0.968225 & 0.968438 & 0.968285 & 0.968440 \\
		\hline
		0.25 & 0.964905 & 0.968072 & 0.967681 & 0.968131 & 0.967898 & 0.968136 & 0.967964 & 0.968137 \\
		\hline
		0.5 & 0.964326 & 0.967737 & 0.967303 & 0.967804 & 0.967542 & 0.967810 & 0.967615 & 0.967811 \\
		\hline
		0.98 & 0.962983 & 0.967002 & 0.966454 & 0.967093 & 0.96675 & 0.967102 & 0.966843 & 0.967104 \\
		\hline
	\end{tabular}
	\label{ta-1}
\end{table}
\begin{table}[htbp]
	\centering
	\setlength\tabcolsep{6pt}
	\caption{The orbital angular momentum $L$ for the periodic orbits with different values of $(z, w, v)$ and the deviation parameters $\alpha_0$. The energy is fixed as $E = 0.96$.}
	\begin{tabular}{|c|c|c|c|c|c|c|c|c|c|c|c|c|}
		\hline
		$\alpha_0$ & $L_{(1,1,0)}$ & $L_{(1,2,0)}$ & $L_{(2,1,1)}$ & $L_{(2,2,1)}$ & $L_{(3,1,2)}$ & $L_{(3,2,2)}$ & $L_{(4,1,3)}$ & $L_{(4,2,3)}$\\
		\hline
		0 & 3.683588 & 3.653406 & 3.657596 & 3.652701 & 3.655335 & 3.652636 & 3.654621 & 3.652616 \\
		\hline
		0.25 & 3.674857 & 3.643534 & 3.647953 & 3.642777 & 3.645578 & 3.642707 & 3.644825 & 3.642684 \\
		\hline
		0.5 & 3.665709 & 3.633064 & 3.637756 & 3.632241 & 3.635245 & 3.632163 & 3.634444 & 3.632138 \\
		\hline
		0.98 & 3.646789 & 3.610898 & 3.616304 & 3.609895 & 3.613446 & 3.609796 & 3.612521 & 3.609763 \\
		\hline
	\end{tabular}
	\label{ta-2}
\end{table}
To support the subsequent analysis of periodic orbits around RBHs with varying deviation parameters $\alpha_0$, we present the required energies and angular momenta for different sets of $(z, w, v)$. The numerical results are provided in Table~\ref{ta-1} and Table~\ref{ta-2}. In Table~\ref{ta-1}, where the angular momentum is fixed at $L = (L_{BMO} + L_\text{ISCO})/2$, we observe that for a given orbit, such as $(2, 1, 1)$, the energy  decreases as $\alpha_0$ increases. This indicates that quantum gravity effects allow particles to maintain periodic orbits with lower energies. Similarly, in Table~\ref{ta-2}, with energy fixed at $E = 0.96$, the required angular momentum $L$ is found to decrease as $\alpha_0$ increases. This further supports the conclusion that quantum gravitational corrections reduce the angular momentum needed for stable orbits near the BH.

\begin{figure}[htbp]
	\centering
	\includegraphics[width=1\textwidth]{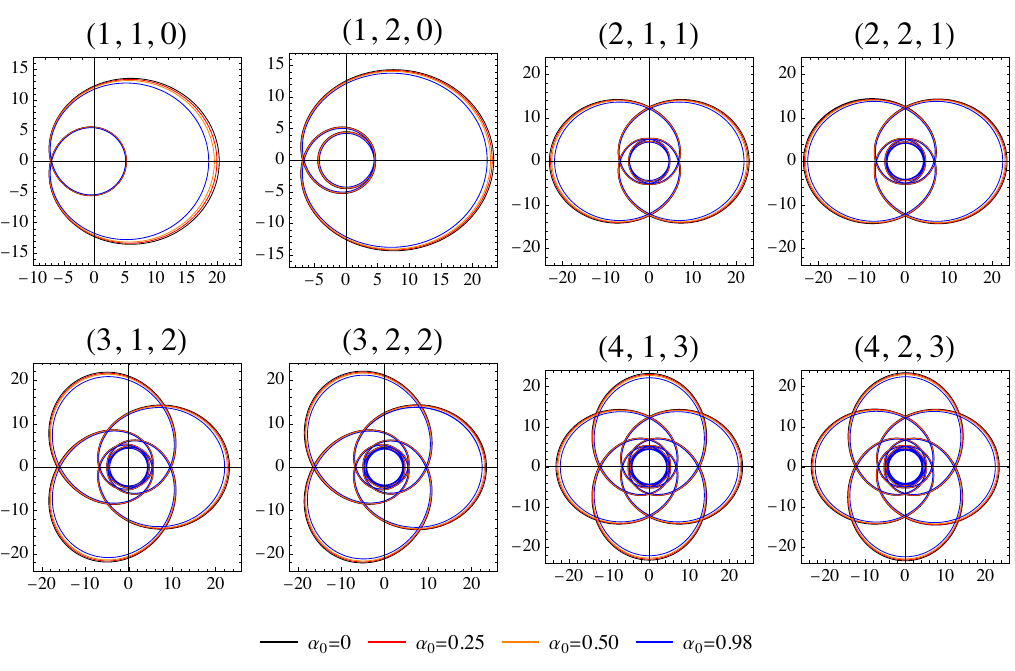}\hspace{4mm}
	\caption{Periodic orbits of different $(z, w, v)$ around RBHs with various deviation parameters $\alpha_0$. Here, we set the  orbital angular momentum $L = (L_\text{MBO} + L_\text{ISCO})/2$ and the corresponding energy $E$ is listed in Table~\ref{ta-1}. }
	\label{orbitchangeafixL}
\end{figure}
\begin{figure}[htbp]
	\centering
	\includegraphics[width=1\textwidth]{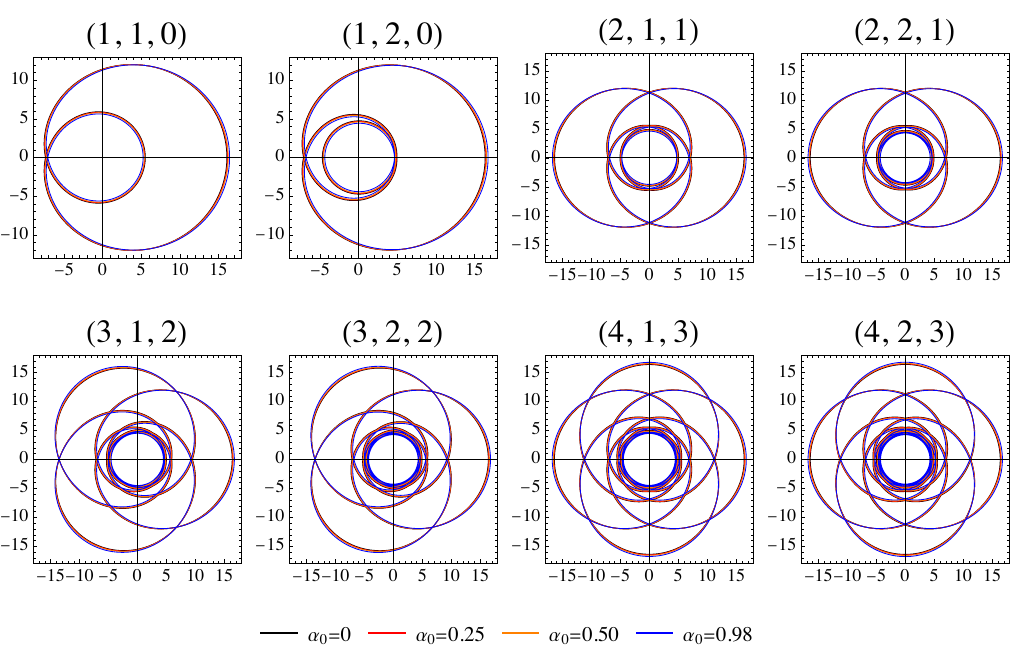}\hspace{4mm}
	\caption{Periodic orbits of different $(z, w, v)$ around RBHs with various deviation parameters $\alpha_0$. Here, we set the energy $E=0.96$ and the corresponding orbital angular momentum $L$ is listed in Table~\ref{ta-2}. }
	\label{orbitchangea}
\end{figure}

Next, we present the periodic orbits with different $(z, w, v)$ for various deviation parameters $\alpha_0$ in Fig.~\ref{orbitchangeafixL} and Fig.~\ref{orbitchangea}. The angular momentum $L$ is fixed in Fig.~\ref{orbitchangeafixL}, while the energy $E$ is kept constant in Fig.~\ref{orbitchangea}. It is evident that periodic orbits with larger zoom number $z$ exhibit more intricate geometric structures, whereas orbits with greater whirl number $w$ complete more revolutions around the central BH between successive apoapses. Moreover, different $\alpha_0$ lead to observable changes in the particle's orbits.

\section{Numerical kludge gravitational waveforms from periodic orbits}\label{sec:NKwaveforms}

\subsection{Waveform modeling via the numerical kludge method}
In the previous section, we have established the existence conditions and dynamical properties of periodic orbits in the RBH spacetime. For further research, we now aim to construct GW signals from these periodic orbits. In EMRI systems, the smaller compact object (secondary) can be regarded as a test particle with mass much smaller than that of the central BH (primary). In the weak-field approximation \mbox{\cite{Hughes:2005qb, Sundararajan:2007jg, Grossman:2011im}}, the energy and angular momentum of the test particle are approximately conserved, and its motion can be modeled as a geodesic orbiting the primary. Therefore, the periodic orbits discussed above can be regarded as a special class of cases in EMRI systems, which is one of the most important  sources for space-based GW detectors.

To accurately model the GWs from these periodic orbits, we adopt the numerical kludge (NK) approach. This method preserves the accuracy of the orbital dynamics while approximating the radiation using  multipole expansion, which  serves as an efficient approximation that captures the essential features of the waveform while losing little accuracy relative to  relativistic Teukolsky waveforms \cite{Babak:2006uv}. Concretely, the first step is to map the particle's trajectory from the Boyer–Lindquist coordinates into the Cartesian coordinates:
\begin{eqnarray}
	x '= r \sin \theta \cos \phi,~~~~y' = r \sin \theta \sin \phi,~~~~z' = r \cos \theta.
\end{eqnarray}
In the weak-field linear approximation, the leading-order gravitational radiation arises from the mass quadrupole moment, which is defined as
\begin{eqnarray}\label{quadrupolemoment}
	I^{i j}=\int  x'^i x'^j T^{t t}\left(t', x'^i\right) d^3x',
\end{eqnarray}
where $T^{tt}$ is the energy density, and $x'^i$ denote the Cartesian coordinates of the secondary. For a point-like particle with mass $m$, the energy density is given by  \cite{Babak:2006uv}:
\begin{eqnarray}\label{stress-energy tensor}
	T^{t t}\left(t', x'^i\right)=m \delta^{(3)}\left(x^i-x'^i(t)\right) .
\end{eqnarray}
where $x^i$ denotes the spatial position of the observer. After integration, the mass quadrupole moment $I^{ij}$ takes a simple form, namely $I^{ij}(t) = m x'^i(t) x'^j(t)$. According to post-Newtonian theory \cite{Blanchet:2013haa}, the metric perturbation $h_{ij}$ in the weak-field limit is proportional to the second-order derivative of the quadrupole moment \cite{peters1964gravitational}:
\begin{eqnarray}\label{GW}
	h_{ij}=\frac{2}{D_{\mathrm{L}}}\frac{d^2I_{ij}}{dt'^2}=\frac{2m}{D_{\mathrm{L}}}(a'_ix'_j+a'_jx'_i+2v'_iv'_j),
\end{eqnarray}
where $D_\text{L}$ is the luminosity distance from the source to the observer, while  $v'_i$ and $a'_i$ denote the velocity and acceleration components of the secondary, respectively. 

To obtain observable gravitational waveforms, we adopt the transverse–traceless (TT) gauge \cite{Fan:2020zhy} and project $h_{ij}$ onto the two polarization components orthogonal to the direction of propagation. The two polarization components $h_+$ and $h_\times$ are defined as \begin{eqnarray}\label{polarizations}
	\begin{aligned}
		h_+&=(h_{\Theta \Theta}-h_{\Phi\Phi})/2,\\
		h_\times&=h_{\Theta\Phi},
	\end{aligned}
\end{eqnarray}
where $h_{\Theta\Theta}$, $h_{\Theta\Phi}$, and $h_{\Phi\Phi}$ are the components of $h_{ij}$ projected into spherical coordinates, and are given by the following expressions:
\begin{eqnarray}\label{component}
	\begin{aligned}
		h_{\Theta \Theta}=&\cos ^2 \Theta\left[h_{x x} \cos ^2 \Phi+h_{x y} \sin 2 \Phi+h_{y y} \sin ^2 \Phi\right]\\
		&+h_{z z} \sin ^2 \Theta-\sin 2 \Theta\left[h_{x z} \cos \Phi+h_{y z} \sin \Phi\right], \\
		h_{\Phi \Theta}=&\cos \Theta\left[-\frac{1}{2} h_{x x} \sin 2 \Phi+h_{x y} \cos 2 \Phi+\frac{1}{2} h_{y y} \sin 2 \Phi\right]\\
		&+\sin \Theta\left[h_{x z} \sin \Phi-h_{y z} \cos \Phi\right], \\
		h_{\Phi \Phi}=&h_{x x} \sin ^2 \Phi-h_{x y} \sin 2 \Phi+h_{y y} \cos ^2 \Phi .
	\end{aligned}
\end{eqnarray}
Here $(\Theta, \Phi)$ represent the latitude and azimuth of the observer, respectively.
Through these computations, we directly connect the periodic orbital motion to the corresponding gravitational waveforms $h_+$ and $h_\times$.

\subsection{Analysis of waveforms without radiation reaction}\label{subsec:noRR}

\begin{figure}[htbp]
	\centering
	\includegraphics[width=1\textwidth]{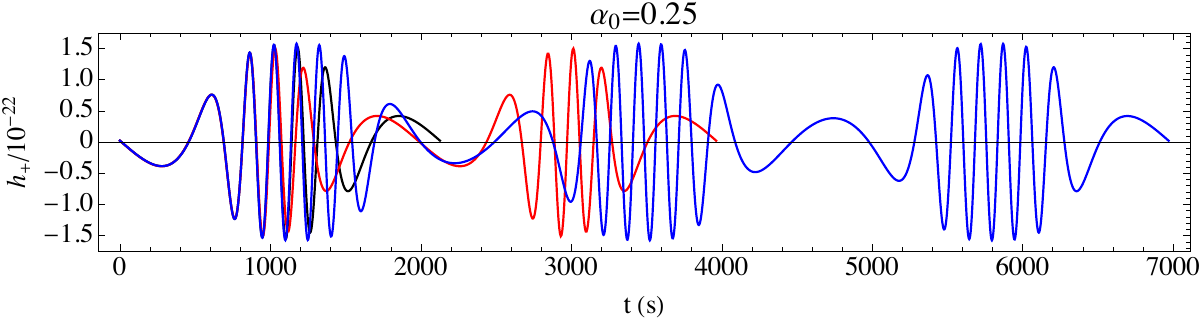}\hspace{4mm}
	\includegraphics[width=1\textwidth]{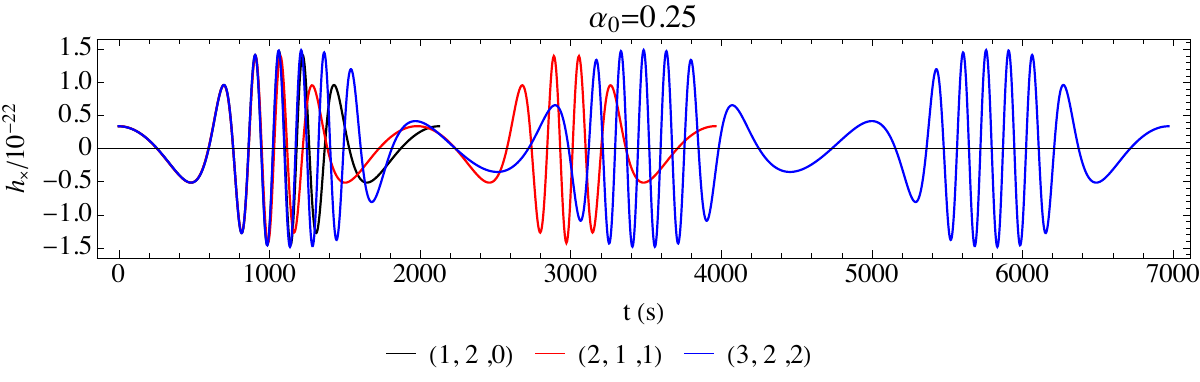}\vspace{0.4mm}
	\caption{GW polarisations $h_{+}$ (upper panel) and $h_{\times}$ (lower panel) for various periodic orbits. Here, we set the deviation parameter $\alpha_{0}=0.25$ and the energy $E=0.96$. The vertical axes are scaled in units of $10^{-22}$.}
	\label{hvsta0p25}
\end{figure}

Having established the connection between  orbits and EMRI gravitational radiation, we can now utilize GW observations to infer the properties of the underlying periodic orbital motion. We set the parameters as follows: the mass of central BH  is $M = 10^6 M_{\odot}$, while the mass of secondary is $m = 10 M_{\odot}$. The system is located at a luminosity distance of $D_L = 2 \text{Gpc}$, and the viewing angles are set to $\Theta = \pi/4$ and $\Phi = \pi/4$, respectively. For details on numerical errors, please refer to the App.~\ref{erroranalysis}. In Fig.~\ref{hvsta0p25}, we present the GW polarizations $h_+$ and $h_\times$ corresponding to several periodic orbits with different sets of $(z, w, v)$. The resulting waveforms exhibit typical zoom and whirl features within one orbital period, reflecting  the complexity of the orbits of the small object around the RBH. It is evident that as the zoom number increases, the waveform develops more intricate substructures, highlighting the sensitivity of the GW radiation to the orbital configuration.

\begin{figure}[htbp]
	\centering
	\includegraphics[width=1\textwidth]{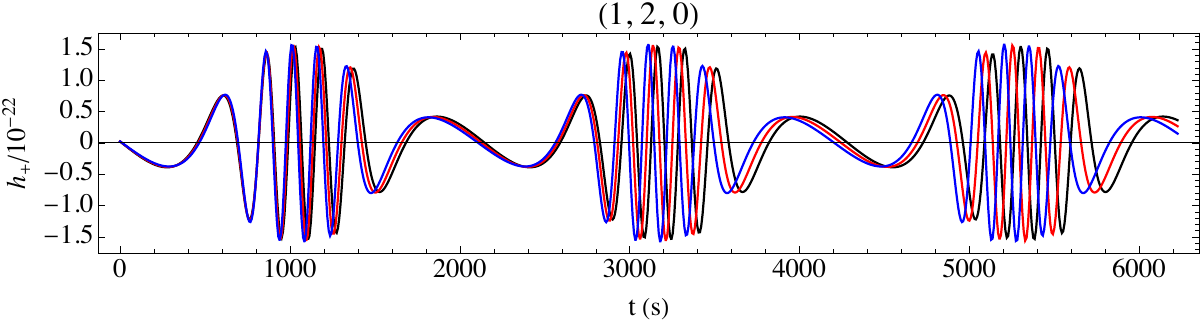}\hspace{4mm}
	\includegraphics[width=1\textwidth]{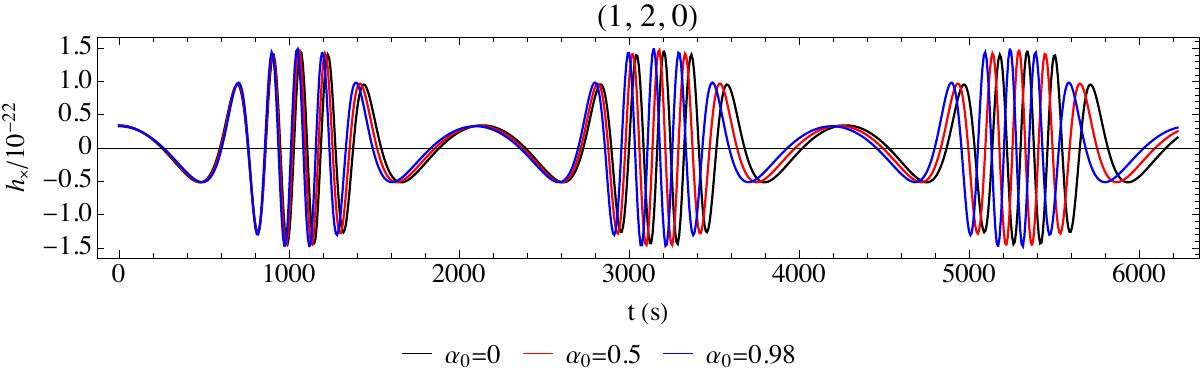}\vspace{0.4mm}
	\caption{GW polarisations $h_{+}$ (upper panel) and $h_{\times}$ (lower panel) for various deviation parameters. Here, we set  $(z, w, v)=(1, 2, 0)$ and the energy $E=0.96$. The vertical axes are scaled in units of $10^{-22}$.}
	\label{hvst}
\end{figure}

To further investigate the effects of quantum gravity on GW, we fix the orbital configuration to $(1, 2, 0)$ and plot the corresponding waveforms for different  $\alpha_0$  in Fig.~\ref{hvst}. We find that as $\alpha_0$ increases, the waveform exhibits a phase advance, and this phase shift accumulates progressively during the orbital evolution, eventually leading to an observable deviation. Meanwhile, the amplitude of the waveform also increases with larger $\alpha_0$, although the overall change remains relatively small. These cumulative deviations suggest that the quantum gravity effects to the background geometry leave measurable imprints on both the phase and amplitude of the gravitational waveform. The sensitivity of the phase shift to the parameter $\alpha_0$ indicates GW signals generated by EMRI systems based on periodic orbits have the potential to serve as precise probes for testing quantum gravity theories.

To quantify differences between two EMRI waveforms $h_a(t)$ and $h_b(t)$, we use the LISA noise-weighted inner product
\begin{equation}
	\langle h_a|h_b\rangle
	= 2 \int_{0}^{\infty} \frac{\tilde h_a^*(f)\tilde h_b(f)+\tilde h_a(f)\tilde h_b^*(f)}{S_n(f)}\,df ,
\end{equation}
where tildes denote Fourier transforms and $S_n(f)$ is the one-sided noise power spectral density~\cite{LISA:2017pwj}.
The normalized faithfulness is
\begin{equation}
	\mathcal{F}(h_a|h_b)=\frac{\langle h_a|h_b\rangle}
	{\sqrt{\langle h_a|h_a\rangle\,\langle h_b|h_b\rangle}} ,
\end{equation}
with $\mathcal{F}=1$ for identical waveforms and $\mathcal{F}=0$ when uncorrelated. As pointed out in Ref.~\cite{Chatziioannou:2017tdw}, two signals can be effectively distinguished by LISA when $\mathcal{F}\leq0.988$.

\begin{figure}[htbp]
	\centering
	\includegraphics[width=0.5\textwidth]{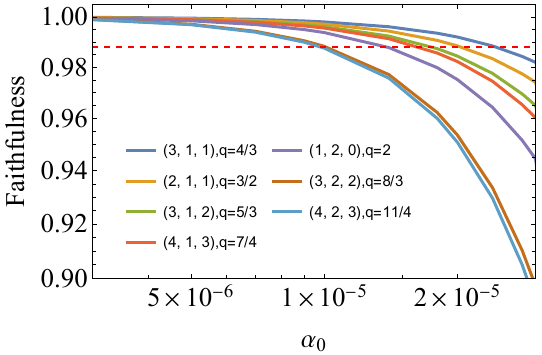}
	\includegraphics[width=0.48\textwidth]{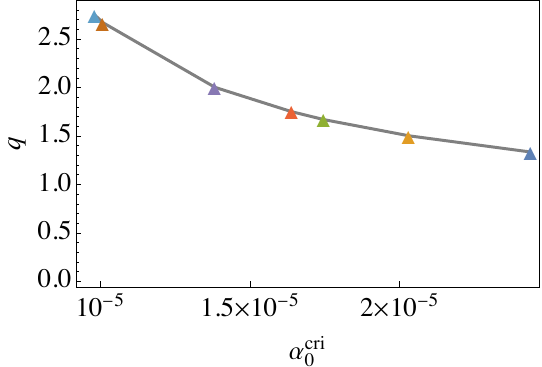}
	\caption{Left panel: The faithfulness of the GW signal between the Schwarzschild BH and  RBH for different  $(z, w, v)$. The red dashed line represents the faithfulness $\mathcal F=0.988$. Right panel: The rational number $q$ as a function of the critical deviation parameter value $\alpha_{0}^{cri}$. The triangles of different colors in the right panel represent the values of the intersections between the corresponding colored curves and the red dashed line in the left panel.
	}
	\label{Faithfulness}
\end{figure}

Finally, we compute the faithfulness between EMRI waveforms from Schwarzschild and RBHs for different orbits, as shown in Fig.~\ref{Faithfulness}. Here, we consider the $h_{+}$ mode of the EMRI waveforms. In the left panel of Fig.~\ref{Faithfulness}, we fix the observation time to one year and discuss the relation between the faithfulness of EMRI waveforms and the deviation parameter $\alpha_0$. The red dashed line in the left panel represents the faithfulness $\mathcal{F}=0.988$, which indicates the detection threshold for one-year period. It is evident that as $\alpha_0$ increases, the overlap between the waveforms decreases, showing that the two BHs become more distinguishable. This phenomenon can be easily understood, as it mainly originates from our above discussion that the deviation parameter $\alpha_0$ affects the evolution of orbits, thereby leading to differences in the waveforms. In addition, we find that for the same $\alpha_0$, as the rational number $q$ increases, the faithfulness correspondingly decreases, suggesting that the two BHs are easier to distinguish. This implies that, if the true EMRI orbit corresponds more closely to a regime with larger $q$, the ability to probe quantum gravity with periodic orbits would be significantly enhanced. In the right panel of Fig.~\ref{Faithfulness}, we present the rational number $q$ as a function of the critical deviation parameter value $\alpha_{0}^{cri}$. The critical deviation parameter $\alpha_{0}^{cri}$ represents the values of the intersections between the corresponding colored curves and the red dashed line in the left panel. Interestingly, we observe that the larger the rational number $q$ is, the smaller the value of $\alpha_0$ required to distinguish the two BHs. This indicates that for larger $q$, the differences between the two BHs can be detected more easily, as even a smaller $\alpha_0$ is sufficient to reveal their distinction. In other words, the deviations between the two BHs become more apparent when $q$ is large.

\subsection{Effects of radiation reaction on orbits and gravitational waveforms}
The waveform analysis presented above is based on the idealized assumption that radiation reaction (RR) is neglected. In this approximation, the orbital energy and angular momentum remain conserved, and the orbits maintain strictly periodic. However, in realistic EMRI systems, the dominant source of gravitational radiation arises from time-varying multipole moments. The resulting continuous energy and angular momentum loss via GWs breaks the periodicity of the orbit over long timescales: the secondary object undergoes a slow secular inspiral and eventually plunges into the central BH. Since waveform phasing accumulates over many orbital cycles, incorporating RR effects is essential for accurate modeling of long-term EMRI evolution.

To quantify the impact of RR, we start with the trace-free mass quadrupole moment, the dominant source of GW radiation in the weak-field regime, defined as
\begin{equation}
	Q_{ij}(t) \equiv I_{ij}(t) - \frac{1}{3}\delta_{ij} I_{k}{}^{k}(t),
\end{equation}
where $I_{ij}(t)$ denotes the standard mass quadrupole moment given in Eq.~\eqref{quadrupolemoment}. At quadrupole-order, the energy and angular momentum fluxes emitted via GWs are then given by~\cite{Mino:2007ft,Blanchet:2013haa,Battista:2021rlh,Battista:2022sci,DeFalco:2024ojf,Ashoorioon:2025ezk}
\begin{equation}\label{ELloss}
	\frac{dE}{dt}\bigg|_{\rm GW}
	= -\frac{1}{5}\,\dddot{Q}_{ij}\dddot{Q}^{ij},
	\qquad
	\frac{dL_i}{dt}\bigg|_{\rm GW}
	= -\frac{2}{5}\,\epsilon_{ijk}\,\ddot{Q}^{j}_{m}\dddot{Q}^{km}.
\end{equation}
These expressions provide a direct and computationally efficient means to estimate radiative losses for a given periodic orbit, explicitly linking GW emission to the secular loss of orbital energy and angular momentum in our framework.

\begin{figure}[htbp]
	\centering
	\begin{minipage}[c]{0.35\textwidth}
		\centering
		\includegraphics[width=\textwidth]{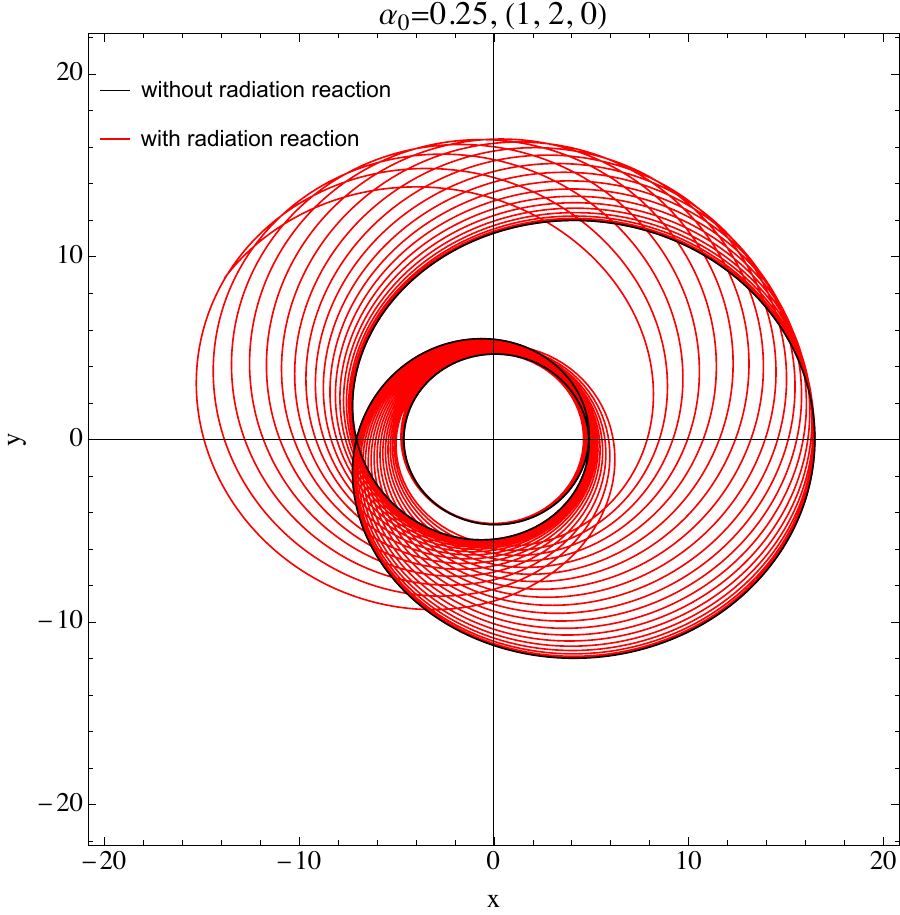}\\
	\end{minipage}
	\hspace{0.03\textwidth} 
	\begin{minipage}[c]{0.55\textwidth}
		\centering
		\includegraphics[width=\textwidth]{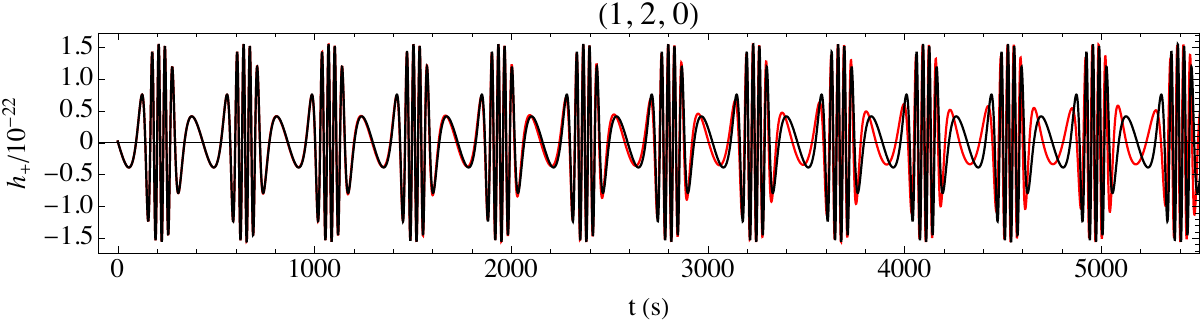}\\
		\vspace{2mm}
		\includegraphics[width=\textwidth]{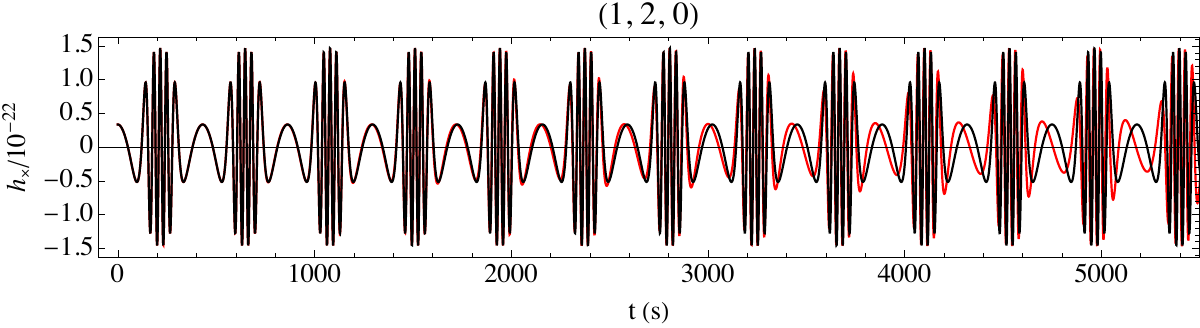}\\
	\end{minipage}
	\caption{Comparison of the orbits (left panel) and waveforms (upper right panel for $h_+$, lower right panel for $h_\times$) with and without RR in the novel RBH model. The parameters are set as follows: $a=0.25$, $(z, w, v)=(1, 2, 0)$, and the initial energy $E=0.96$. The vertical axes are scaled in units of $10^{-22}$.}
	\label{RRLingnRRLingv0}
\end{figure}

\begin{figure}[htbp]
	\centering
	\begin{minipage}[c]{0.35\textwidth}
		\centering
		\includegraphics[width=\textwidth]{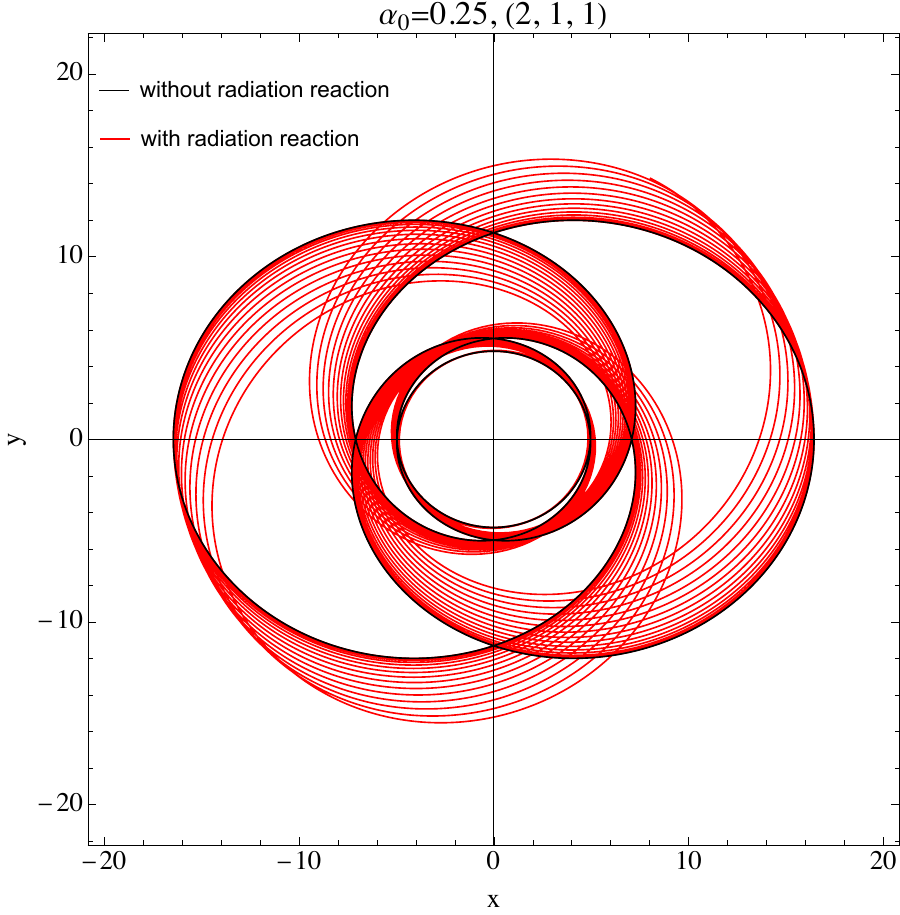}\\
	\end{minipage}
	\hspace{0.03\textwidth} 
	\begin{minipage}[c]{0.55\textwidth}
		\centering
		\includegraphics[width=\textwidth]{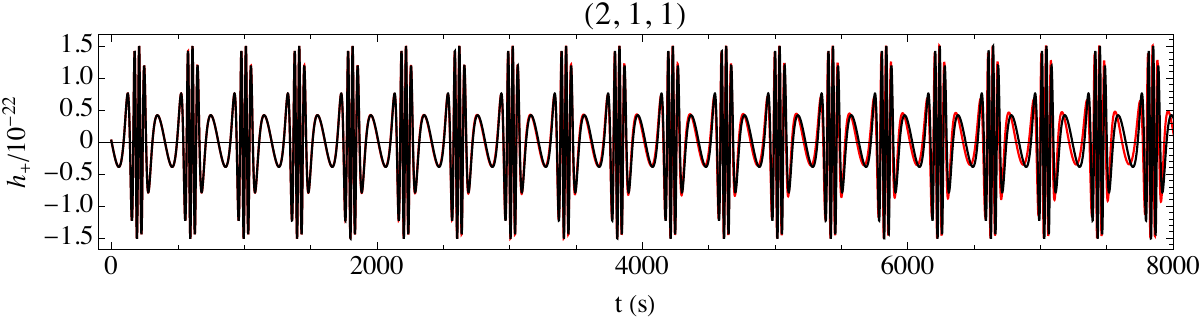}\\
		\vspace{2mm}
		\includegraphics[width=\textwidth]{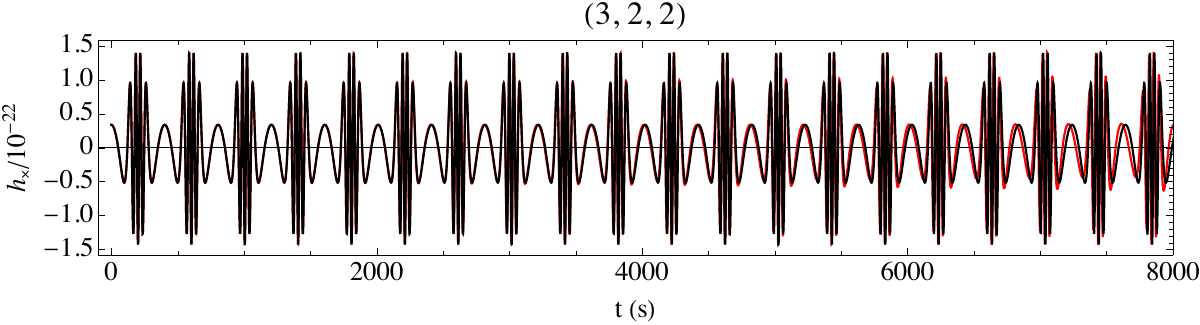}\\
	\end{minipage}
	\caption{Comparison of the orbits (left panel) and waveforms (upper right panel for $h_+$, lower right panel for $h_\times$) with and without RR in the novel RBH model. The parameters are set as follows: $a=0.25$, $(z, w, v)=(2, 1, 1)$, and the initial energy $E=0.96$. The vertical axes are scaled in units of $10^{-22}$.}
	\label{RRLingnRRLingv1}
\end{figure}

\begin{figure}[htbp]
	\centering
	\begin{minipage}[c]{0.35\textwidth}
		\centering
		\includegraphics[width=\textwidth]{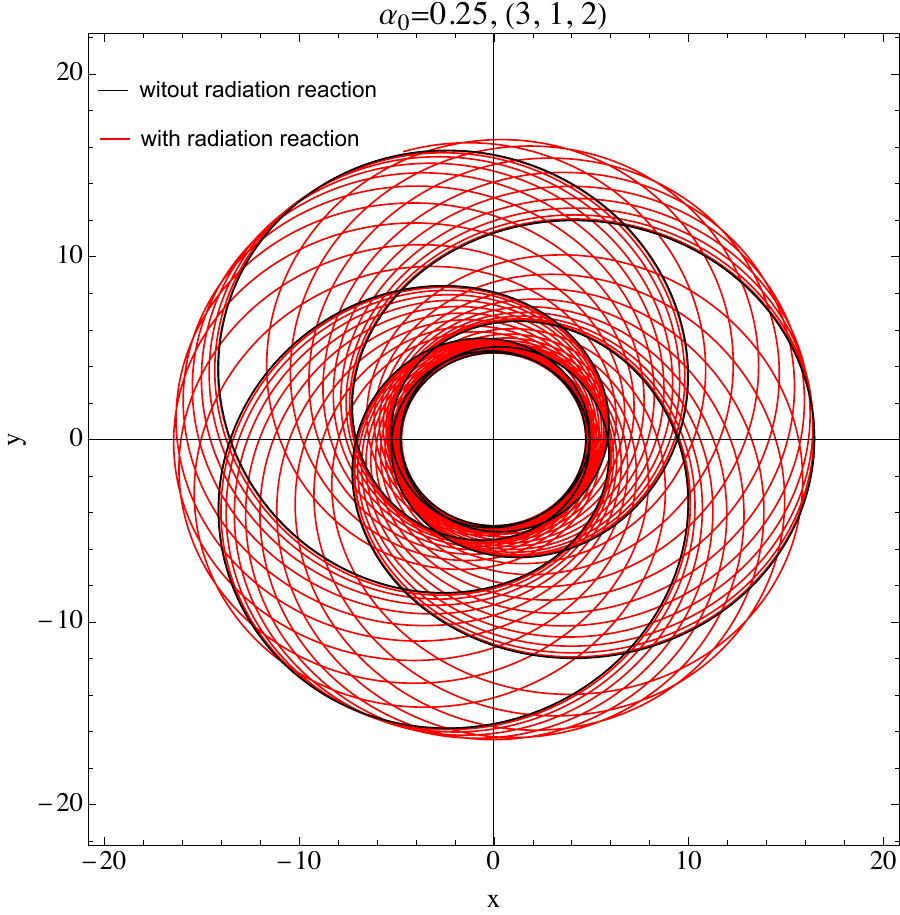}\\
	\end{minipage}
	\hspace{0.03\textwidth} 
	\begin{minipage}[c]{0.55\textwidth}
		\centering
		\includegraphics[width=\textwidth]{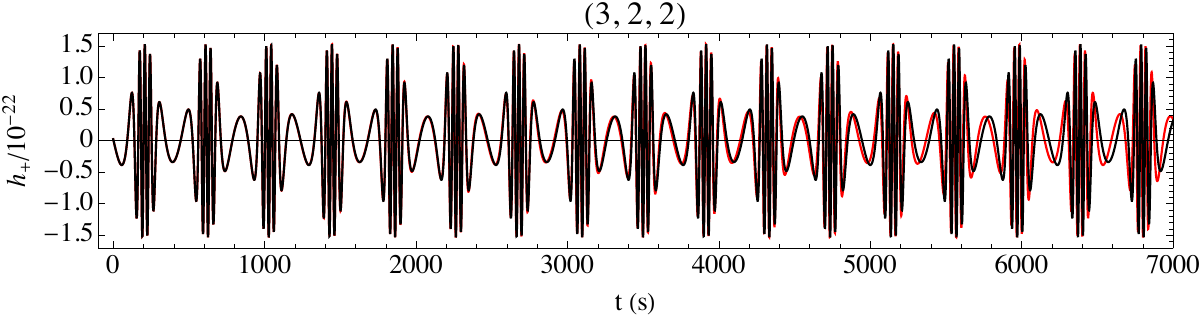}\\
		\vspace{2mm}
		\includegraphics[width=\textwidth]{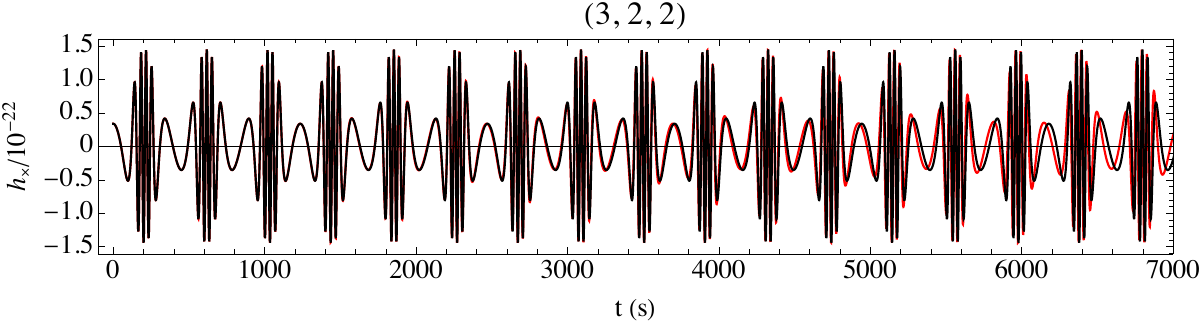}\\
	\end{minipage}
	\caption{Comparison of the orbits (left panel) and waveforms (upper right panel for $h_+$, lower right panel for $h_\times$) with and without RR in the novel RBH model. The parameters are set as follows: $a=0.25$, $(z, w, v)=(3, 1, 2)$, and the initial energy $E=0.96$. The vertical axes are scaled in units of $10^{-22}$.}
	\label{RRLingnRRLingv2}
\end{figure}

\begin{figure}[htbp]
	\centering
	\includegraphics[width=0.5\textwidth]{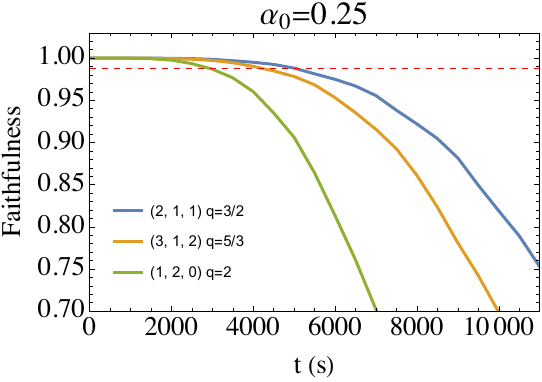}
	\caption{Faithfulness of the $h_+$ with/without RR for different $(z, w, v)$ in the RBH model. Here, we fix $\alpha_{0}=0.25$. The red dashed line represents the faithfulness $\mathcal F=0.988$.}
	\label{Faithfulnessv1}
\end{figure}

Figs.~\ref{RRLingnRRLingv0}-\ref{RRLingnRRLingv2} compare the orbits and gravitational waveforms with and without RR at fixed parameters for several periodic orbits in our RBHs model. The key differences are summarized as follows:
\begin{itemize}
	\item As shown in the left panel of each figure in Figs.~\ref{RRLingnRRLingv0}-\ref{RRLingnRRLingv2}, the orbit computed without RR (solid black line) remains strictly periodic, retracing the same trajectory over successive cycles. In contrast, when RR is included (solid red line), the orbit undergoes an evident long-term evolution, where the angular precession induced by RR causes the orbit to deviate from its original trajectory, thereby breaking its periodicity.
	
	\item The corresponding GW waveforms are modified accordingly (as shown in right panels of each figure in Figs.~\ref{RRLingnRRLingv0}-\ref{RRLingnRRLingv2}). For both polarization modes $h_+$ and $h_\times$, the waveforms that include RR exhibit a cumulative phase advance relative to those without RR. This advance occurs because the inspiral slightly shortens each orbital period. Furthermore, a slight increase in the GW amplitude can be observed over time.
\end{itemize}

To quantify the impact of RR, we further calculate the faithfulness between waveforms with and without RR based on the analysis in Sec.~\ref{subsec:noRR}. Fig.~\ref{Faithfulnessv1} shows the faithfulness of the $h_+$ polarization for different periodic orbits $(z, w, v)$ in the RBH model with $\alpha_0 = 0.25$. It can be seen that over the finite duration of orbital evolution considered, the faithfulness drops below the threshold at which the two waveforms can be reliably distinguished. This result underscores the necessity of incorporating RR effects in realistic EMRI waveform modeling, which is essential for probing quantum gravity effects. Moreover, we observe that a larger rational number $q$ leads to a shorter time required to distinguish between the two types of waveforms. This finding is consistent with the conclusion drawn in Sec.~\ref{subsec:noRR}.

\section{A comparative study of modified BH models: novel RBHs, Hayward BHs and LQG-corrected BHs}\label{sec:reular-Hay-LQG}
The issue of singularities in BH has motivated the proposal of various modified models. Among them, the novel RBH with Minkowski core, Hayward BH~\cite{Hayward:2005gi}, and qOS BH~\cite{Lewandowski:2022zce} are representative schemes. Although these models share a similar asymptotic behavior at large scale, they differ significantly in their spacetime structure and the physical origin of the corrections. This section compares their orbital dynamics and GW characteristics, aiming to reveal observable differences arising from different modification mechanisms.

For completeness, we summarize the explicit forms of their metric functions and highlight the connections between them:
\begin{itemize}
	\item \textbf{The novel RBHs} (with parameters $x=1$ and $c=3$): At large scales ($r\to\infty$), its metric function expands as
	\[
	f_{\text{RBH}}(r)\cong 1-\frac{2M}{r}\left(1-\frac{\alpha_{0}M}{r^3}+\cdots\right).
	\]
	\item \textbf{The Hayward BHs}: The metric function is given by
	\[
	f_{\text{Hayward}}(r)=1-\frac{2Mr^2}{r^3+M \alpha_{0}},
	\]
	whose expansion at large scales coincides exactly with that of the novel RBH:
	\[
	f_{\text{Hayward}}(r)\cong 1-\frac{2M}{r}\left(1-\frac{\alpha_{0}M}{r^3}+\cdots\right).
	\]
	\item \textbf{The qOS BHs}: The metric function derived in Ref.~\cite{Lewandowski:2022zce} reads
	\[
	f_{\text{qOS}}(r)=1-\frac{2M}{r}+\frac{2\alpha_{0}M^2}{r^4}=1-\frac{2M}{r}\left(1-\frac{\alpha_{0}M}{r^3}\right),
	\]
	which precisely reproduces the asymptotic behavior of the two previous models.
\end{itemize}

\begin{figure}[htbp]
	\centering
	\includegraphics[width=0.48\textwidth]{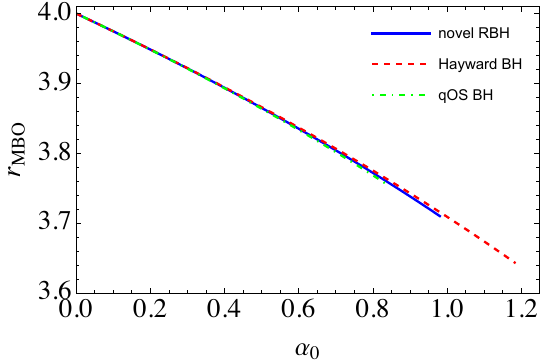}\hspace{4mm}
	\includegraphics[width=0.48\textwidth]{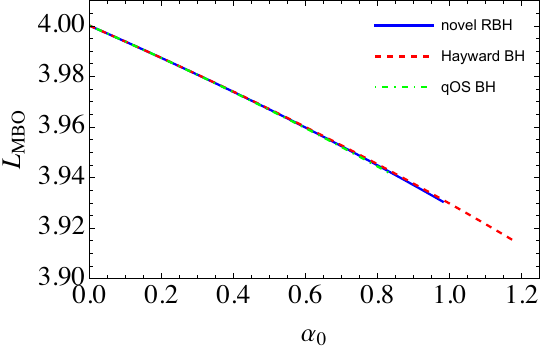}\vspace{0.4mm}
	\caption{The radius $r_\text{MBO}$ (left panel) and the angular momentum $L_\text{MBO}$ (right panel) as a function of the deviation parameter $\alpha_{0}$ for the three BH models.}
	\label{MBOvsa03BH}
\end{figure}

\begin{figure}[htbp]
	\centering
	\includegraphics[width=0.48\textwidth]{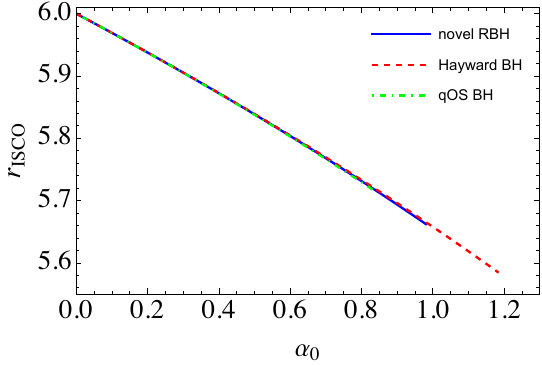}\hspace{4mm}
	\includegraphics[width=0.48\textwidth]{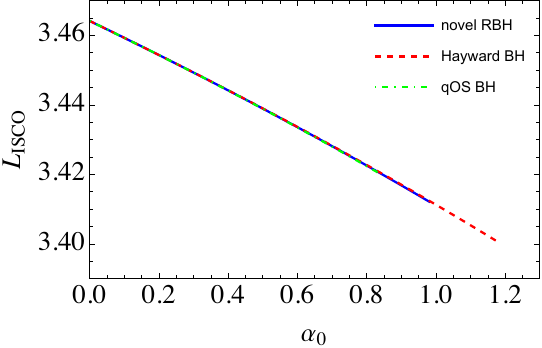}\vspace{0.4mm}
	\includegraphics[width=0.48\textwidth]{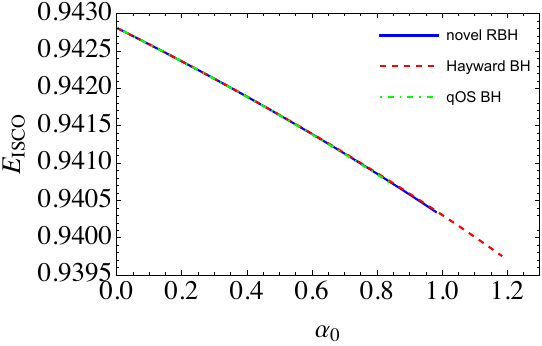}\vspace{0.4mm}
	\caption{The radius $r_\text{ISCO}$ (upper left panel), the angular momentum $L_\text{ISCO}$ (upper right panel), and the energy $E_\text{ISCO}$ (lower panel) as functions of the deviation parameter $\alpha_{0}$ for for the three BH models.}
	\label{ISCOvsa03BH}
\end{figure}

Fig.~\ref{MBOvsa03BH} shows the dependence of the MBO radius (left panel) and angular momentum (right panel) on the deviation parameter $\alpha_0$ for the three BH models. In all cases, both quantities decrease monotonically as $\alpha_0$ increases. For a fixed $\alpha_0$, the differences among the three models are small, yet a clear ordering is discernible: the qOS BH yields the smallest values, the novel RBH lies in between, and the Hayward BH gives the largest. A similar trend is observed for the ISCO radius, angular momentum and energy in Fig.~\ref{ISCOvsa03BH}. These results indicate that the three BH models exhibit very similar behaviors, with the differences becoming even smaller for relatively small values of $\alpha_0$.

\begin{figure}[htbp]
	\centering
	\includegraphics[width=0.6\textwidth]{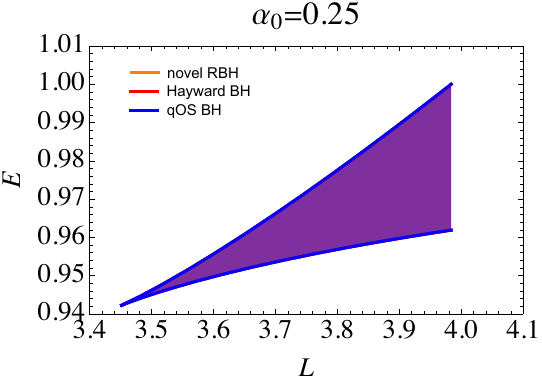}\hspace{4mm}	
	\caption{The allowed $(E,L)$ regions for bound orbits of massive particles around different black holes, with the deviation parameter fixed at $\alpha_{0}=0.25$.}
	\label{LErange3BH}
\end{figure}

\begin{figure}[htbp]
	\centering
	\includegraphics[width=0.5\textwidth]{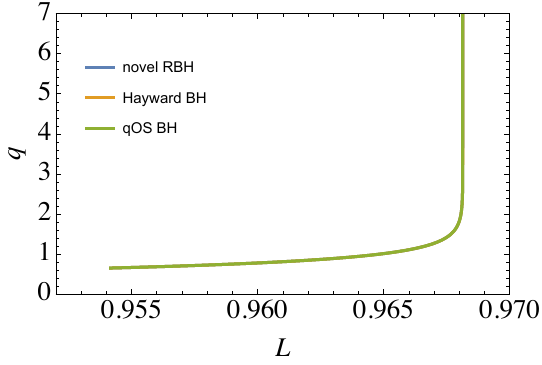}\vspace{0.4mm}
	\includegraphics[width=0.47\textwidth]{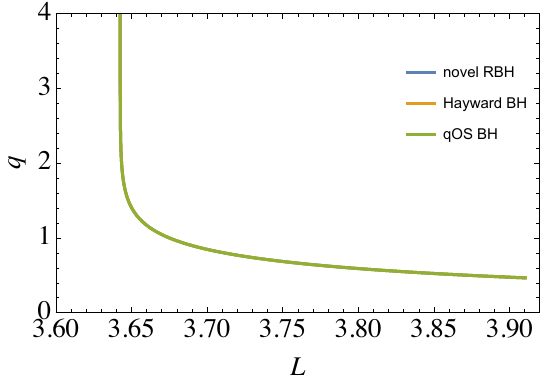}\vspace{0.4mm}
	\caption{Left panel: The rational number $q$ as a function of the energy of periodic orbits $E$ for different BHs with $\alpha_{0}=0.25$, at fixed angular momentum $L = (L_\text{MBO} + L_\text{ISCO})/2$. Right panel: The rational number $q$ as a function of the angular momentum of periodic orbits $L$ for different BHs with $\alpha_{0}=0.25$, at fixed energy $E = 0.96$.}
	\label{qvsEandLv0}
\end{figure}

\begin{figure}[htbp]
	\centering
	\includegraphics[width=1\textwidth]{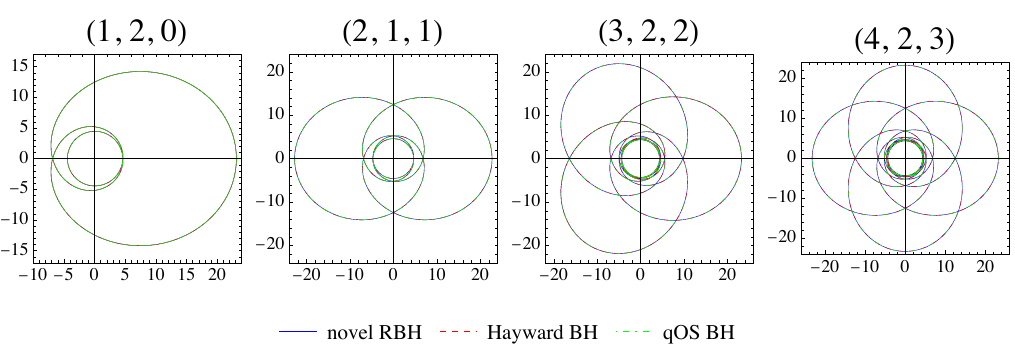}\hspace{4mm}
	\caption{Periodic orbits for various $(z, w, v)$ in different BHs with $\alpha_0=0.25$. Here, we set the  orbital angular momentum $L = (L_\text{MBO} + L_\text{ISCO})/2$ and the corresponding energy $E$ is listed in Table~\ref{ta-3BH-1}. }
	\label{orbit3BHfixLL}
\end{figure}

\begin{figure}[htbp]
	\centering
	\includegraphics[width=1\textwidth]{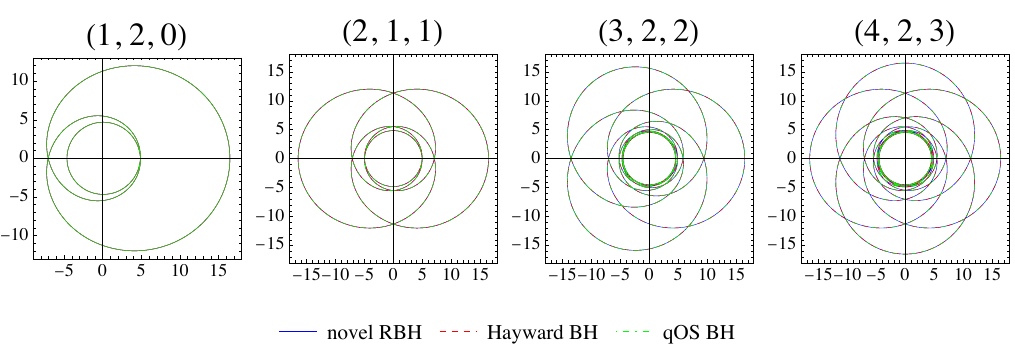}\hspace{4mm}
	\caption{Periodic orbits for various $(z, w, v)$ in different BHs with $\alpha_0=0.25$. Here, we set the energy $E=0.96$ and the corresponding orbital angular momentum $L$ is listed in Table~\ref{ta-3BH-2}. }
	\label{orbit3BHfixEE}
\end{figure}

\begin{figure}[htbp]
	\centering
	\includegraphics[width=1\textwidth]{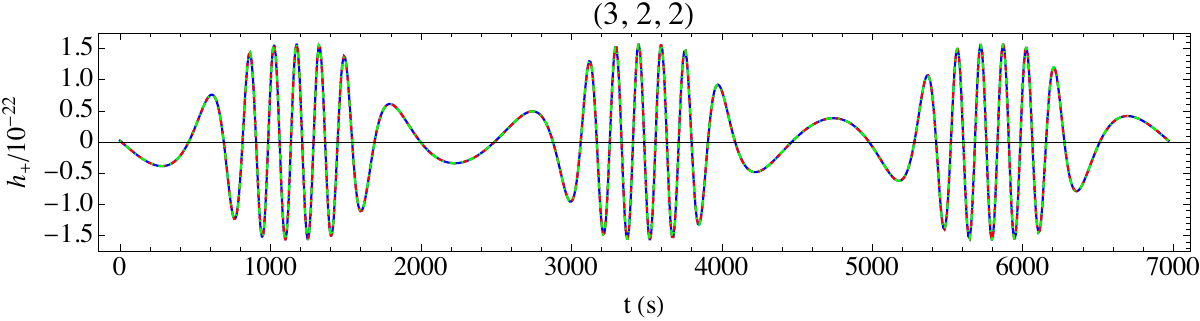}\hspace{4mm}
	\includegraphics[width=1\textwidth]{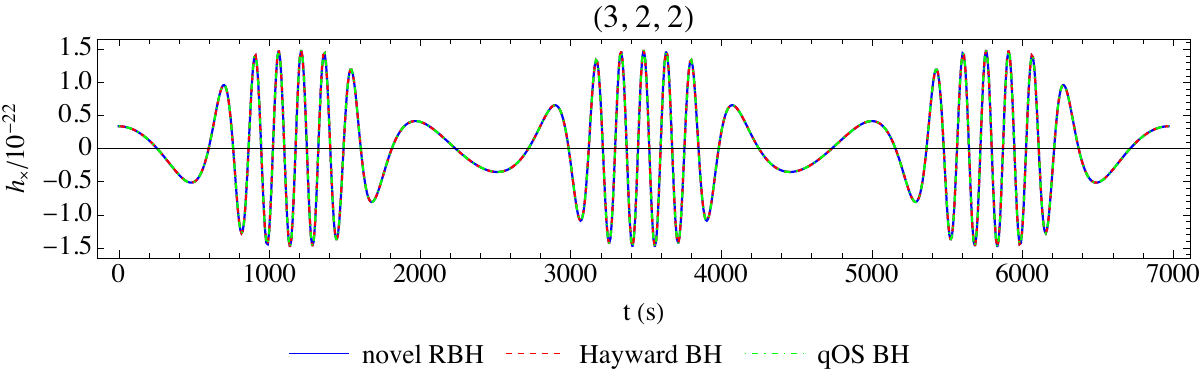}\vspace{0.4mm}
	\caption{GW polarisations $h_{+}$ (upper panel) and $h_{\times}$ (lower panel) for various BHs. Here, we set $(z, w, v)=(3, 2, 2)$, the deviation parameter $\alpha_{0}=0.25$ and the energy $E=0.96$. The vertical axes are scaled in units of $10^{-22}$.}
	\label{hvst3BHa0p25}
\end{figure}

We now focus on a moderate value, $\alpha_0=0.25$, to examine the differences among the three models in more detail\footnote{We have checked other values of $\alpha_0$ and obtained qualitatively similar conclusions.}. Figs. \ref{LErange3BH}-\ref{qvsEandLv0} display the allowed $(E,L)$ regions for bound orbits and the dependence of the rational number $q$ on $E$ and $L$, respectively. It can be clearly observed that the behavior of $q$ is almost identical across the three models, suggesting that their periodic orbits may be difficult to distinguish directly. To verify this observation and quantify differences, we select several representative triplets $(z, w, v)$ and plot the corresponding periodic orbits (Fig. \ref{orbit3BHfixLL}-\ref{orbit3BHfixEE}) and GW waveforms (Fig. \ref{hvst3BHa0p25}), with the relevant numerical data listed in Tables~\ref{ta-3BH-1} and~\ref{ta-3BH-2}. Both the geometric shapes of the orbits and the resulting waveforms are remarkably similar, making them practically indistinguishable. Furthermore, a closer inspection of the tables reveals that, for a fixed $q$, the corresponding orbital energy and angular momentum values are nearly identical across the three models, confirming quantitatively that the periodic orbits around these BHs are macroscopically indistinguishable.

The underlying reason for this near-indistinguishability lies in the consistency of the large-scale spacetime structure among the three models. In the large-scale region far from the BH center, all three metric functions share exactly the same leading and sub-leading terms. This renders the spacetime geometry in that region highly similar. Meanwhile, the existence interval of periodic orbits is strictly confined between the MBO and the ISCO, and this region maintains a certain radial distance from the BH center. Therefore, the inherent differences in the near-horizon structures of the three types of BHs are significantly weakened in this interval, ultimately leading to the macroscopic geometric morphologies of the periodic orbits being difficult to distinguish.

\begin{table}[htbp]
	\centering
	\setlength\tabcolsep{6pt}
	\caption{The energy $E$ for the periodic orbits with different values of $(z, w, v)$ in different BHs. Here, we fix $\alpha_0=0.25$ and $L = (L_\text{MBO} + L_\text{ISCO})/2$.}
	\begin{tabular}{|c|c|c|c|c|c|c|c|c|c|c|c|c|}
		\hline
		$\text{BH}$ & $E_{(1,2,0)}$ & $E_{(2,1,1)}$ & $E_{(3,2,2)}$ & $E_{(4,2,3)}$\\
		\hline
		Novel RBHs & 0.96807235 & 0.96768092 & 0.96813557 & 0.96813707 \\
		\hline
		Hayward BHs & 0.96807294 & 0.96768168 & 0.96813612 & 0.96813761 \\
		\hline
		qOS BHs & 0.96807176 & 0.96768015 & 0.96813502 & 0.96813652 \\
		\hline
	\end{tabular}
	\label{ta-3BH-1}
\end{table}
\begin{table}[htbp]
	\centering
	\setlength\tabcolsep{6pt}
	\caption{The orbital angular momentum $L$ for the periodic orbits with different values of $(z, w, v)$ in different BHs . Here, we fix $\alpha_0=0.25$ and $E = 0.96$.}
	\begin{tabular}{|c|c|c|c|c|c|c|c|c|c|c|c|c|}
		\hline
		$\text{BH}$ & $L_{(1,2,0)}$ & $L_{(2,1,1)}$ & $L_{(3,2,2)}$ & $L_{(4,2,3)}$\\
		\hline
		Novel RBHs & 3.64353443 & 3.64795267 & 3.64270671 & 3.64268428\\
		\hline
		Hayward BHs & 3.64354820 & 3.64796516 & 3.64272086 & 3.64269846 \\
		\hline
		qOS BHs & 3.64352060 & 3.64794014 & 3.64269249 & 3.64267006 \\
		\hline
	\end{tabular}
	\label{ta-3BH-2}
\end{table}

\section{Conclusion and discussion}\label{sec:conclusion}

In this work, we have investigated the dynamics of test particles on periodic orbits around a spherically symmetric RBH with a Minkowski core, and the associated GW emission from EMRIs. We have also supplemented a comparative study of the novel RBH with the Hayward BH and the qOS BH to reveal the similarities and differences induced by different modification mechanisms.

Firstly, by analyzing the effective potential and the admissible energy-angular momentum range, we explored how the deviation parameter $\alpha_0$ modifies orbits. Taking the MBO and ISCO as fiducial points, we identified the region where periodic orbits can exist. Our results show that increasing $\alpha_0$ drives both MBO and ISCO to smaller radii and reduces their  angular momentum. The energy of the ISCO monotonically decreases with $\alpha_0$ while the MBO remains at the maximum energy $E=1$. These trends arise from the modified effective potential, indicating that quantum gravity effects allow particles to remain bound closer to the BH with reduced angular momentum and critical energy. Furthermore, the $(E, L)$ parameter space for bound orbits is reshaped compared to the Schwarzschild case, reflecting the global impact of $\alpha_0$ on orbital dynamics.

Secondly, focusing on periodic orbits, we employed the rational-number classification scheme $(z, w, v)$ to characterize their geometry. Our analysis shows that both the orbital energy and angular momentum required to sustain such orbits decrease monotonically with increasing $\alpha_0$. This indicates that quantum gravity effects reduce the critical energy for the bound motion near the BH. In addition, larger values of $z$ or $w$ lead to more intricate orbital structures, exhibiting the characteristic ``zoom-whirl" behavior. These structural differences, which vary with $\alpha_0$, further indicate that deviations from Schwarzschild geometry leave observable imprints on the periodic orbits, offering a potential avenue to probe the nature of RBHs through GW observations.

Thirdly, we have extended our investigation of periodic orbits around spherically symmetric RBHs to their associated GW emissions in the context of EMRI systems. Modeling the secondary object as a test particle, we employed the NK method to compute the corresponding gravitational waveforms $h_+$ and $h_\times$. The resulting waveforms capture characteristic features of the underlying periodic motion, including ``zoom-whirl" behavior associated with the orbital integers $(z, w, v)$. We found that increasing the deviation parameter $\alpha_0$ leads to detectable phase shifts and slight amplitude variations in the GW signal, indicating that deviations from the Schwarzschild geometry leave observable imprints on the waveform. When RR effects are considered, the periodicity of the orbit is broken due to the loss of energy and angular momentum, leading to inspiral behavior. This further induces cumulative phase advances and amplitude increases in the GW waveforms, and the faithfulness analysis shows that the differences between waveforms with and without RR become distinguishable over time.

Our analysis of waveform faithfulness further indicates a clear positive dependence of the  distinguishability between Schwarzschild and RBHs on both the deviation parameter $\alpha_0$ and the rational number $q$. In particular, for larger $q$, even small values of $\alpha_0$ suffice to produce noticeable differences, thereby enhancing the detectability of departures from Schwarzschild geometry. These results suggest that GWs from EMRI systems based on periodic orbits offer a promising avenue to probe quantum gravity effects encoded in RBH metrics.

Moreover, the comparative study of the novel RBH, Hayward BH and qOS BH shows that the three models share the same large-scale asymptotic behavior of the metric function, leading to highly similar periodic orbit geometries and GW waveforms in the bound orbit region between MBOs and ISCOs. Quantitative analysis reveals that under the same $\alpha_0$, the qOS BH has the smallest MBOs and ISCOs radius and angular momentum, followed by the novel RBH, while the Hayward BH has the largest. This difference originates from the distinct spacetime structures near the horizon of the three BHs, but this difference is weakened in the bound orbit region far from the center, resulting in macroscopic indistinguishability of periodic orbits and waveforms.

Looking ahead, this work can be extended in several promising directions. Future studies may investigate rotating RBH and generalize the analysis to spinning test particles. Furthermore, the applicability of the proposed rational-number classification scheme should be tested in different background spacetimes to explore its validity and universality within broader theoretical frameworks.

\appendix

\section{The influence of the parameter $c$}\label{ceffect}

In the metric of the novel RBH, the exponent $c$ appearing in the potential term $P(r)=-(M/r)\exp{(-\alpha_{0}M^x/r^c)}$ plays a crucial role in governing the radial falloff of the gravitational potential\footnote{The dimensionless exponent $x$ governs the mass-dependent strength of the quantum-gravity correction term $\alpha_0 M^x/r^c$. Consequently, it regulates the mass scaling of the Kretschmann scalar maximum $K_{\rm max}$ \cite{Ling:2021olm}. In this paper we set $M=1$, so varying $x$ does not affect any quantitative results.}. This section systematically investigates the influence of $c$ on the spacetime geometry, orbital dynamics, and GW waveforms, highlighting its distinct effect from that of $\alpha_0$.
\begin{figure}[htbp]
	\centering
	\includegraphics[width=0.48\textwidth]{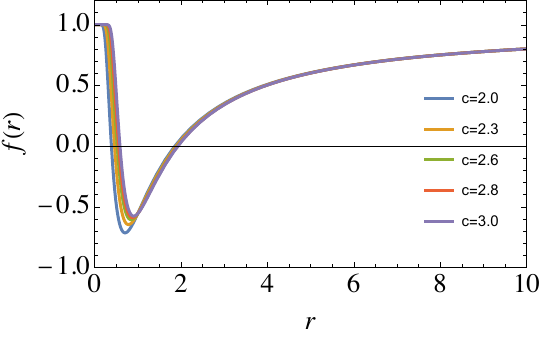}\hspace{4mm}
	\caption{The metric function $f(r)$ for various values of the parameter $c$, with fixed $\alpha_{0}=0.25$ and $x=1$.}
	\label{FigfrandK1}
\end{figure}

To ensure the curvature remains sub-Planckian, the parameters must satisfy the constraints $c \geq x \geq c/3$ and $c \geq 2$ \cite{Ling:2021olm, Zhang:2024nny}. In this appendix, we fix $\alpha = 0.25$ and $x = 1$, which restricts $c$ to the range $2 \leq c \leq 3$. Fig.~\ref{FigfrandK1} illustrates how the metric function $f(r)$ varies with $c$. The results show that as $c$ increases, $f(r)$ undergoes significant changes in the small-$r$ region ($r \ll 1$), while the variation in the large-$r$ region is comparatively mild. This indicates that the quantum corrections are predominantly concentrated near the black hole event horizon.

\begin{figure}[htbp]
	\centering
	\includegraphics[width=0.48\textwidth]{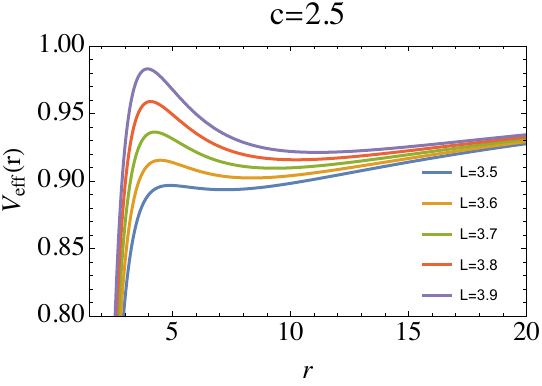}\hspace{4mm}
	\includegraphics[width=0.48\textwidth]{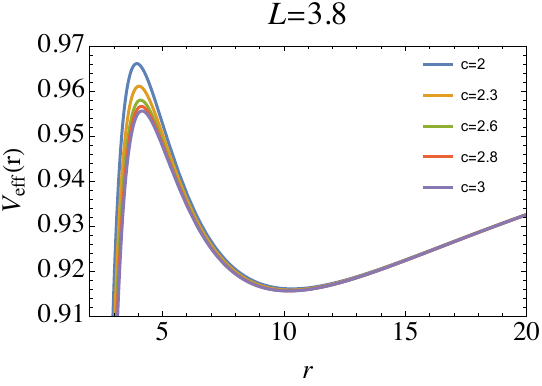}\vspace{0.4mm}
	\caption{Left panel: The effective potential $V_{\text{eff}}$ for different values of the orbital angular momentum $L$, with $c = 2.5$. Right panel: The effective potential $V_{\text{eff}}$ for different values of the parameter $c$, with $L = 3.8$. In both panels, we fix $\alpha_{0} = 0.25$ and $x = 1$.}
	\label{Veffv11}
\end{figure}

Next, we illustrate in Fig.~\ref{Veffv11} the behavior of the effective potential $V_{\rm eff}$ for different orbital angular momentum $L$ and parameter $c$. For a given value of $c$, the local maximum of $V_{\rm eff}$ increases with $L$ (left panel of Fig.~\ref{Veffv11}), consistent with the findings in Sec.~\ref{subsec:ds2}. In contrast, the right panel of Fig.~\ref{Veffv11} demonstrates that for fixed $L$, the local maximum of $V_{\rm eff}$ decreases as $c$ increases.

\begin{figure}[htbp]
	\centering
	\includegraphics[width=0.48\textwidth]{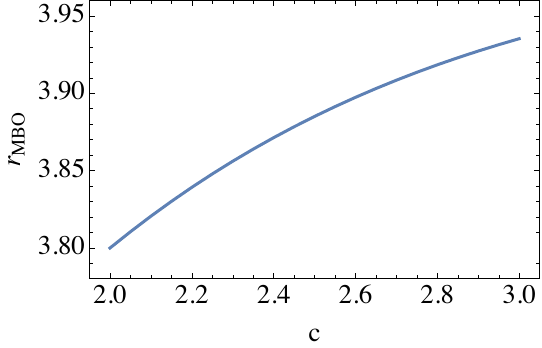}\hspace{4mm}
	\includegraphics[width=0.48\textwidth]{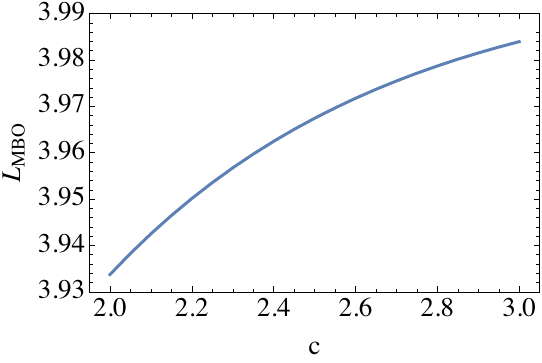}\vspace{0.4mm}
	\caption{The radius $r_\text{MBO}$ (left panel) and the angular momentum $L_\text{MBO}$ (right panel) as a function of the parameter $c$.}
	\label{MBOvsc}
\end{figure}

\begin{figure}[htbp]
	\centering
	\includegraphics[width=0.48\textwidth]{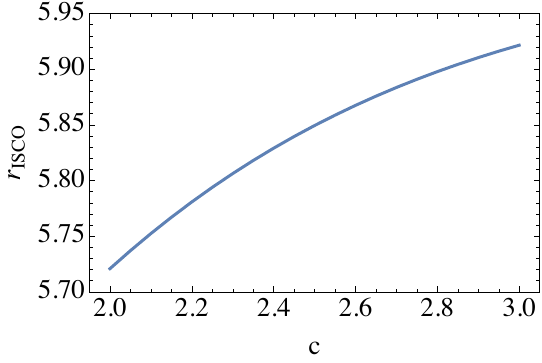}\hspace{4mm}
	\includegraphics[width=0.48\textwidth]{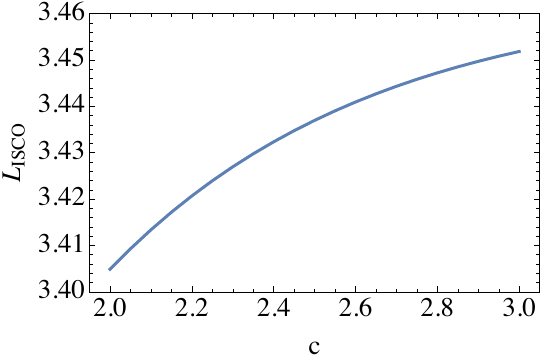}\vspace{0.4mm}
	\includegraphics[width=0.48\textwidth]{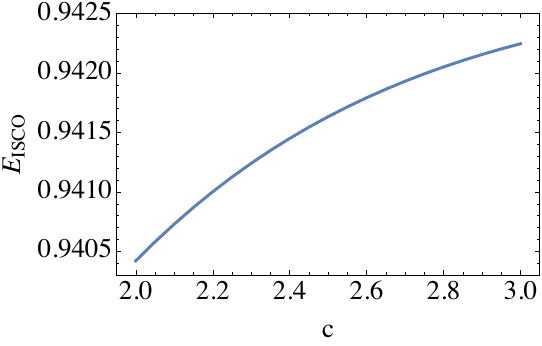}\vspace{0.4mm}
	\caption{The radius $r_\text{ISCO}$ (upper left panel), the angular momentum $L_\text{ISCO}$ (upper right panel) and the energy $E_\text{ISCO}$ (lower panel) as a function of the parameter $c$.}
	\label{ISCOvsc}
\end{figure}

\begin{figure}[htbp]
	\centering
	\includegraphics[width=0.7\textwidth]{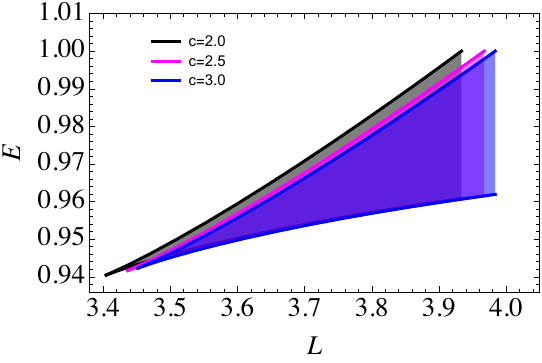}\hspace{4mm}
	\caption{The allowed $(E-L)$ regions for the massive particle’s bound orbits with different parameters $c$.}
	\label{LErange1}
\end{figure}

As shown in Figs.~\ref{MBOvsc}-\ref{LErange1}, the parameter $c$ influences the boundaries of bound orbits in the following ways. First, both the radius and angular momentum of the MBO increase with $c$ (Fig.~\ref{MBOvsc}), a trend opposite to that induced by $\alpha_0$. Second, the radius, angular momentum, and energy of the ISCO also increase with $c$ (Fig.~\ref{ISCOvsc}), indicating that a larger $c$ shifts stable orbits outward. Third, as $c$ increases, the allowed $(E-L)$ region for bound orbits shifts to the right, again opposing the effect of $\alpha_0$ (Fig.~\ref{LErange1}). These results demonstrate that $c$ and $\alpha_0$ exert opposite effects on the orbital boundaries.

\begin{figure}[htbp]
	\centering
	\includegraphics[width=0.48\textwidth]{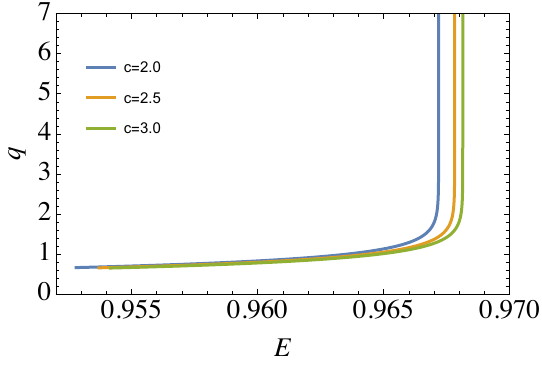}\hspace{4mm}
	\includegraphics[width=0.46\textwidth]{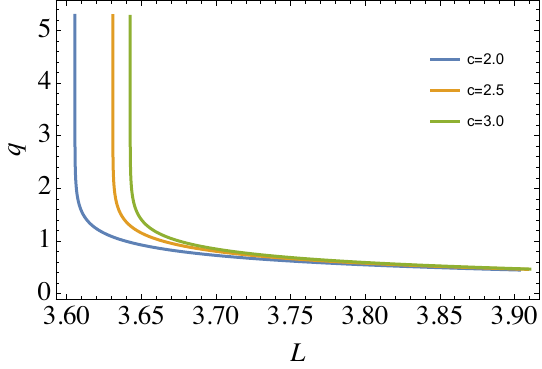}\vspace{0.4mm}
	\caption{Left panel: Rational number $q$ as a function of the energy $E$ for periodic orbits with different $c$, at fixed angular momentum $L=(L_\text{MBO}+L_\text{ISCO})/2$. Right panel: $q$ as a function of $L$ for different $c$, at fixed energy $E=0.96$.}
	\label{qvsEandLv11}
\end{figure}

\begin{figure}[htbp]
	\centering
	\includegraphics[width=1\textwidth]{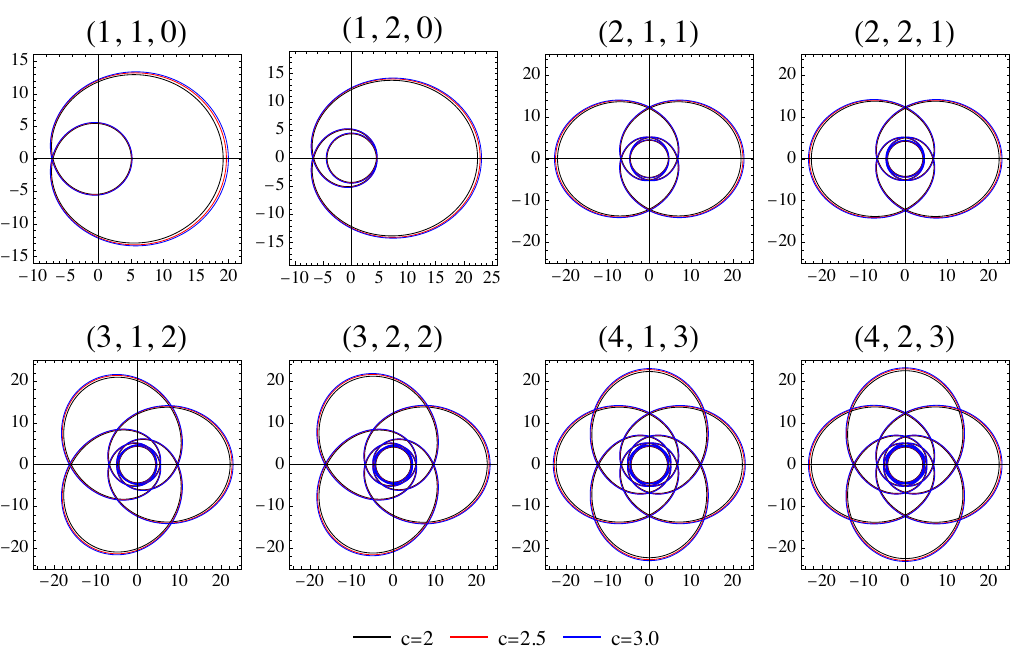}\hspace{4mm}
	\caption{Periodic orbits of different $(z, w, v)$ around RBHs with various  $c$. Here, we set the  orbital angular momentum $L = (L_\text{MBO} + L_\text{ISCO})/2$ and the corresponding energy $E$ is listed in Table~\ref{ta-1}. }
	\label{orbitchangecfixL}
\end{figure}

\begin{figure}[htbp]
	\centering
	\includegraphics[width=1\textwidth]{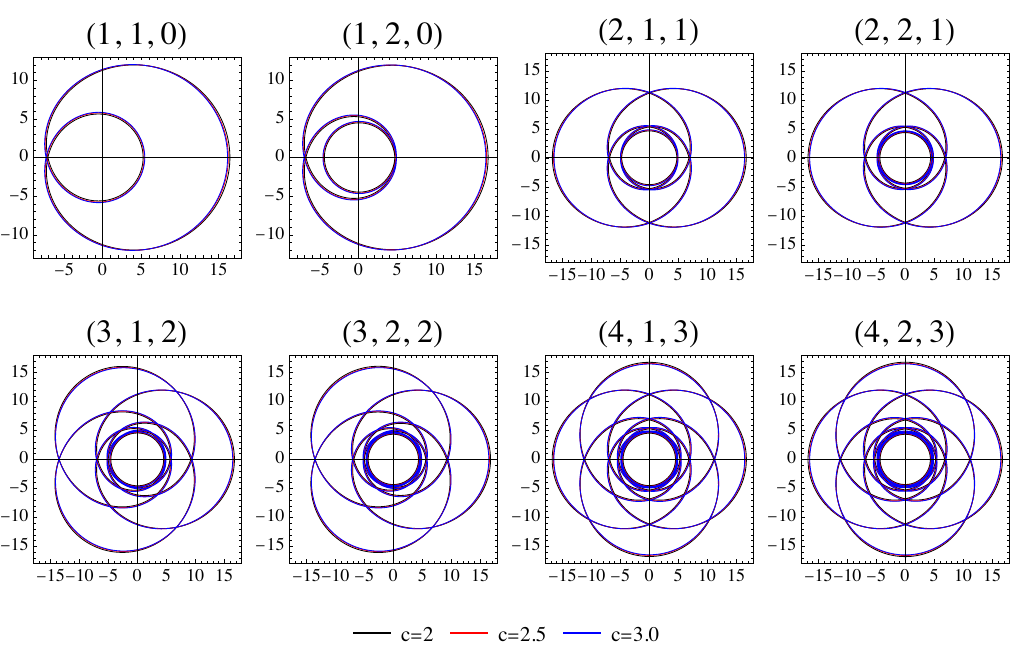}\hspace{4mm}
	\caption{Periodic orbits of different $(z, w, v)$ around RBHs with various $c$. Here, we set the energy $E=0.96$ and the corresponding orbital angular momentum $L$ is listed in Table~\ref{ta-2}. }
	\label{orbitchangecfixE}
\end{figure}

\begin{figure}[htbp]
	\centering
	\includegraphics[width=1\textwidth]{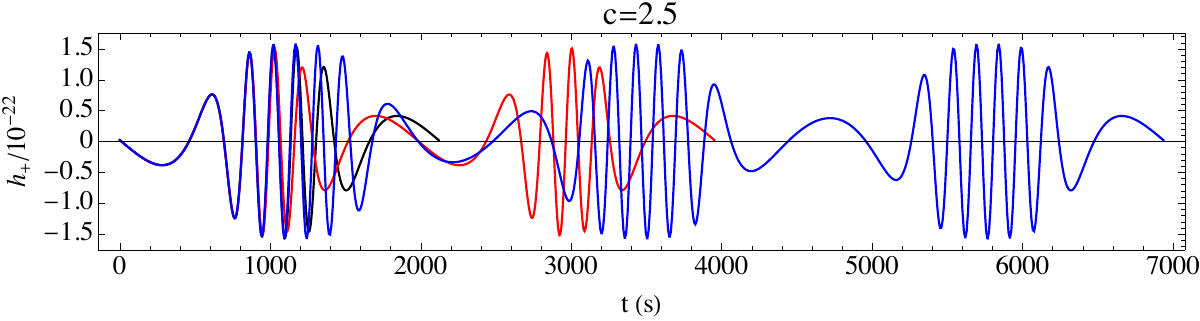}\hspace{4mm}
	\includegraphics[width=1\textwidth]{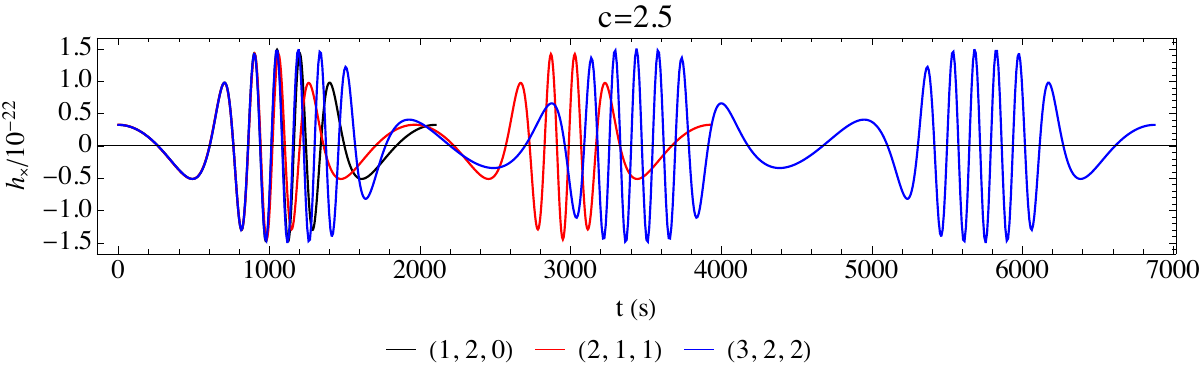}\vspace{0.4mm}
	\caption{GW polarisations $h_{+}$ (upper panel) and $h_{\times}$ (lower panel) for various periodic orbits. Here, we set the parameter $c=2.5$ and the energy $E=0.96$. The vertical axes are scaled in units of $10^{-22}$.}
	\label{hvstc2p5}
\end{figure}

\begin{figure}[htbp]
	\centering
	\includegraphics[width=1\textwidth]{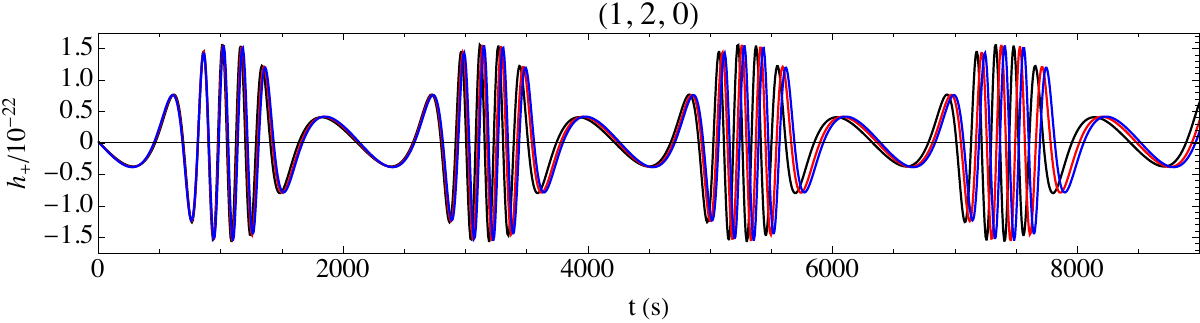}\hspace{4mm}
	\includegraphics[width=1\textwidth]{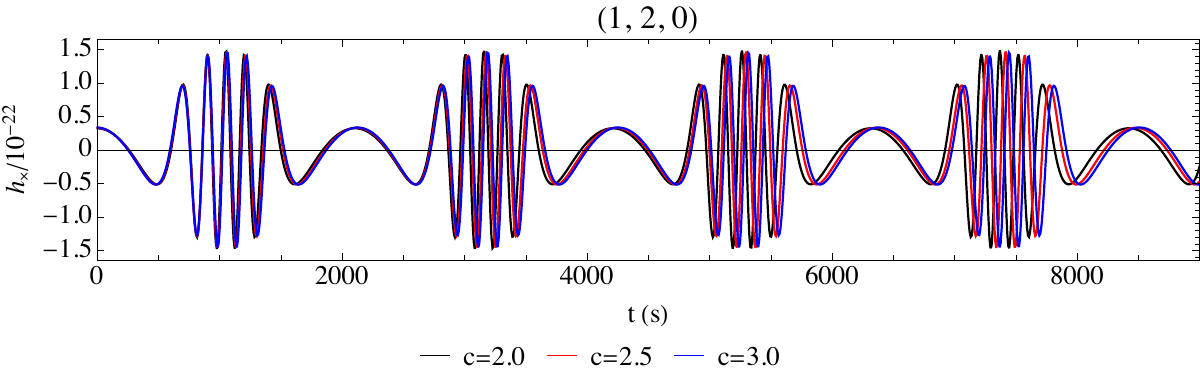}\vspace{0.4mm}
	\caption{GW polarisations $h_{+}$ (upper panel) and $h_{\times}$ (lower panel) for various $c$. Here, we set $(z, w, v)=(1, 2, 0)$ and the energy $E=0.96$. The vertical axes are scaled in units of $10^{-22}$.}
	\label{hvstq120}
\end{figure}

We further illustrate the influence of $c$ on the rational number $q$ in Fig.~\ref{qvsEandLv11}. Variations in $c$ also affect the geometric structure of periodic orbits and, consequently, the associated gravitational waveforms. To demonstrate this, we select a series of triplets $(z, w, v)$ and present the corresponding orbital diagrams (Figs. \ref{orbitchangecfixL}-\ref{orbitchangecfixE}) as well as gravitational waveforms (Figs. \ref{hvstc2p5}-\ref{hvstq120}), with the relevant data summarized in Tables~\ref{ta-3} and~\ref{ta-4}.

In summary, the parameters $c$ and $\alpha_0$ exhibit distinctly different effects on gravitational waveforms. An increase in $\alpha_0$ reduces the radius and angular momentum of both the MBO and ISCO, allowing particles to maintain bound orbits closer to the BH. This is reflected in the waveforms as observable cumulative phase advances, accompanied by a slight increase in amplitude as $\alpha_0$ grows. In contrast, an increase in $c$ leads to larger radii and angular momenta for the MBO and ISCO, shifting stable orbits outward. For orbits with the same $(z, w, v)$ values, this results in waveforms with smaller amplitudes and observable cumulative phase lags.

\begin{table}[htbp]
	\centering
	\setlength\tabcolsep{6pt}
	\caption{Energy $E$ for periodic orbits with different $(z, w, v)$ and $c$, at fixed $L=(L_\text{MBO}+L_\text{ISCO})/2$.}
	\begin{tabular}{|c|c|c|c|c|c|c|c|c|c|c|c|c|}
		\hline
		$c$ & $E_{(1,1,0)}$ & $E_{(1,2,0)}$ & $E_{(2,1,1)}$ & $E_{(2,2,1)}$ & $E_{(3,1,2)}$ & $E_{(3,2,2)}$ & $E_{(4,1,3)}$ & $E_{(4,2,3)}$\\
		\hline
		2 & 0.963614 & 0.967103 & 0.966661 & 0.967171 & 0.966905 & 0.967177 & 0.966979 & 0.967179 \\
		\hline
		2.5 & 0.964444 & 0.967746 & 0.967333 & 0.967808 & 0.967561 & 0.967814 & 0.967631 & 0.967815 \\
		\hline
		3 & 0.964905 & 0.968072 & 0.967681 & 0.968131 & 0.967898 & 0.968136 & 0.967964 & 0.968137 \\
		\hline
	\end{tabular}
	\label{ta-3}
\end{table}
\begin{table}[htbp]
	\centering
	\setlength\tabcolsep{6pt}
	\caption{Angular momentum $L$ for periodic orbits with different $(z, w, v)$ and $c$, at fixed energy $E=0.96$.}
	\begin{tabular}{|c|c|c|c|c|c|c|c|c|c|c|c|c|}
		\hline
		$c$ & $L_{(1,1,0)}$ & $L_{(1,2,0)}$ & $L_{(2,1,1)}$ & $L_{(2,2,1)}$ & $L_{(3,1,2)}$ & $L_{(3,2,2)}$ & $L_{(4,1,3)}$ & $L_{(4,2,3)}$\\
		\hline
		2 & 3.638869 & 3.606556 & 3.611138 & 3.605766  & 3.608678 & 3.605693 & 3.607896 & 3.605669 \\
		\hline
		2.5 & 3.663812 & 3.631943 & 3.636459 & 3.631165 & 3.634034 & 3.631093 & 3.633264 & 3.631070 \\
		\hline
		3 & 3.674857 & 3.643534 & 3.647953 & 3.642777 & 3.645578 & 3.642707 & 3.644825 & 3.642684 \\
		\hline
	\end{tabular}
	\label{ta-4}
\end{table}


\section{Numerical error analysis}\label{erroranalysis}

To ensure the reliability of our numerical results for gravitational waveforms and orbital dynamics, we conduct the error analysis and verify the numerical convergence of our calculations. The primary source of numerical error in this work stems from the integration of the geodesic equations for test particles in the RBH spacetime. We employ a fourth-order Runge-Kutta method with a fixed time step size $h$ to solve the radial and azimuthal equations of motion.

To quantify the numerical errors and confirm convergence, we introduce the relative deviations of the orbital variables as
\begin{eqnarray}\delta t = \left| \frac{t_h - t_{href}}{t_{href}} \right|, \quad \delta r = \left| \frac{r_h - r_{href}}{r_{href}} \right|, \quad \delta \phi = \left| \frac{\phi_h - \phi_{href}}{\phi_{href}} \right|,\end{eqnarray} 
where $t_h$, $r_h$, and $\phi_h$ denote the numerical solutions obtained with an step size $h$, while $t_{href}$, $r_{href}$, $\phi_{href}$ are the reference solutions computed with the step size $h = 1/2$. We test a series of step sizes $h =  1, 2, 4, 8, 16$ and analyze how $\delta t$, $\delta r$, and $\delta \phi$ scale with $h$.

\begin{figure}[htbp]
	\centering
	\includegraphics[width=0.45\textwidth]{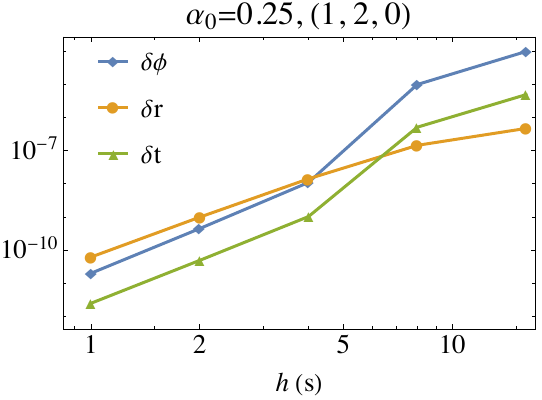}\hspace{4mm}
	\includegraphics[width=0.45\textwidth]{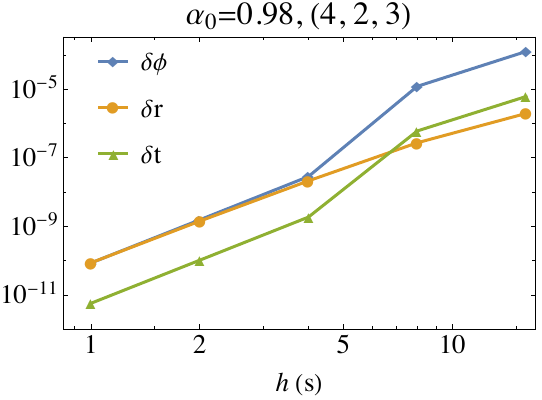}\hspace{4mm}
	\caption{Variation of the orbital deviations $(\delta t, \delta r, \delta \phi)$ with step size $h$.  Left panel:  $\alpha_0=0.25$, $(z,w,v)=(1,2,0)$, $E=0.96$. Right panel: $\alpha_0=0.98$, $(z,w,v)=(4,2,3)$, $E=0.96$. The reference step size is $h=1/2$.}
	\label{ErrorsFig}
\end{figure}

Fig.~\ref{ErrorsFig} shows the logarithmic variation of orbital deviations with the integration step size for two representative periodic orbits. It can be observed that all deviation quantities decrease as the integration step size $h$ is reduced, consistent with the expected convergence order of the fourth-order Runge-Kutta method. For sufficiently small step sizes (e.g., $h \leq 2$), the deviations $\delta r$ and $\delta \phi$ are suppressed to levels below $10^{-8}$, confirming the accuracy of our numerical implementation. In all production runs, we adopt a step size of $h = 1/2$ to ensure high precision, justifying our choice through this convergence test.

In addition, it is important to note that the NK method itself introduces a systematic approximation error due to the use of the quadrupole formula in the weak-field limit. However, previous studies \cite{Babak:2006uv} have demonstrated that NK waveforms achieve a waveform overlap of over 95\% with fully relativistic Teukolsky-based waveforms for EMRI systems, validating the reliability of the NK approach for our analysis.

\section*{Acknowledgments}

We are very grateful to Qin Tan, Guoyang Fu, Shulan Li and Weiliang Qian for helpful discussions and suggestions.We also express our gratitude to the referee, whose insightful comments and suggestions greatly improved this paper. This work is supported by Natural Science Foundation of China (Grants Nos. 12275079, 12447156, 12447137, 12035005, 12547143, and 12375055), China Post-doctoral Science Foundation (Grant No. 2025M773339) and Postgraduate Scientific Research Innovation Project of Hunan Province under Grant Nos. CX20220509 and 2024JJ1006.

\bibliographystyle{style1}
\bibliography{Ref}
\end{document}